\documentclass[review,showpacs,showkeyword,10pt]{elsarticle}
\usepackage{amssymb}
\usepackage{geometry}
\journal{}

\usepackage[english]{babel}
\usepackage{algorithm}
\usepackage{algorithmic}
\usepackage{subfigure}
\usepackage{amsfonts}
\usepackage{amsmath}
\usepackage{enumerate}
\usepackage{framed}
\usepackage{lscape}
\usepackage{multicol}
\usepackage{amsthm}
\usepackage{ae}
\usepackage[T1]{fontenc}
\usepackage[ansinew]{inputenc}
\usepackage{color}
\usepackage[colorlinks]{hyperref}
\usepackage{bm}
\usepackage[mathlines]{lineno}
\usepackage{xspace}
\usepackage{algorithm}
\usepackage{hyperref}
\usepackage{graphicx}% Include figure files
\usepackage{dcolumn}% Align table columns on decimal point
\usepackage{bm}% bold math
\usepackage{hyperref}% add hypertext capabilities
\usepackage[mathlines]{lineno}% Enable numbering of text and display math
\usepackage{multicol}
\usepackage{amssymb}
\usepackage{lineno,hyperref}
\input{epsf}
\newcommand{\figref}[1]{{Figure~\ref{#1}}}

\newcommand{\secref}[1]{{Section~\ref{#1}}}
\renewcommand{\eqref}[1]{{Eq.~(\ref{#1})}}

\newcommand{\sech}{\mbox{\textrm{\ensuremath{\,}sech}}}

\newcommand{\acknowledgement}{\textbf{Acknowledgements\\}}
\biboptions{sort&compress,square}
\begin{document}
\begin{frontmatter}
%%% Title, authors and addresses
\title{\textbf{Nonlinear dynamics of DNA systems with inhomogeneity effects}}
\author[biophys,cetic]{\textbf{J. Brizar Okaly}\corref{cor1}}\ead{okalyjoseph@yahoo.fr}\cortext[cor1]{Corresponding author}
\author[biophys,cetic]{\textbf{Alain Mvogo}} \ead{mvogal\_2009@yahoo.fr}
\author[meca,cetic]{\textbf{R. Laure Woulach\'{e}}} \ead{rwoulach@yahoo.com}
\author[meca,cetic]{\textbf{T. Cr\'{e}pin Kofan\'{e}}} \ead{tckofane@yahoo.com}
\address[biophys]{Laboratory of Biophysics, Department of Physics, Faculty of Science, University of Yaounde I, P.O. Box 812, Yaounde, Cameroon}
\address[meca]{Laboratory of Mechanics, Department of Physics, Faculty of Science, University of Yaounde I, P.O. Box 812, Yaounde, Cameroon}
\address[cetic]{African Centre of Excellence in Information and Communication Technologies, University of Yaounde I, P.O. Box 812, Yaounde, Cameroon.}

\date{\today}
\begin{abstract}
We investigate the nonlinear dynamics of the Peyrard-Bishop DNA model taking into account site dependent inhomogeneities. By means of the multiple-scale expansion in the semi-discrete approximation, the dynamics is governed by the perturbed nonlinear Schr\"{o}dinger equation. We carry out a multiple-scale soliton perturbation analysis to find the effects of the variety of nonlinear inhomogeneities on the breatherlike soliton solution. During the crossing of the inhomogeneities, the coherent structure of the soliton is found stable. The global shape of the inhomogeneous molecule is merged with the shape of the homogeneous molecule. However, the velocity, the wavenumber and the angular frequency undergo a time-dependent correction that is proportional to initial width of the soliton and depends on the nature of the inhomogeneities.
}
\end{abstract}
\end{frontmatter}                      %display desired
%
%\tableofcontents
%%
%\newpage
%
\section{\label{intro} Introduction}
DNA plays an important role in the carrier, protection, transmission, suppression, replication, transcription, recombination, repair and mutation of genetic information in biological systems. Several processes in the cell start with the binding of an protein enzyme at a promoter site of the DNA. This binding is known to change the conformation of DNA by generating a nonlinear localized excitation, which causes the opening of few base pairs in the molecular chain \cite{sty}. This excitation explains the transition conformation, the regulation of transcription, the denaturation and charge transport in terms of polarons and bubbles \cite{kal}. Many theoretical models based on the longitudinal and transversal motions (as well as bending, stretching and rotations) have been proposed to describe the dynamics of DNA double helix \cite{Englander, yak1, yak2, kong, oka2}. One of the most popular of these models is the Peyrard-Bishop (PB) model in which DNA is considered as a helicoidal structure with a collection of particles connected with springs \cite{PB,dau}.

The biological processes which are executed in nature are carried out in a setting of inhomogeneities, such as the protein enzymes which catalyse billion chemical reactions occurring anytime in biological systems. These processes are taking place not as isolated entities, but they do so in a molecular crowded environment. The particular biological functions of DNA imply the presence of different sites along the strands, such as promotor ($P$), coding ($C$),  several regulatory regions, ($R_1$, $R_2$, $R_3$), terminator ($T$), which contain a specific sequence of base pairs and naturally make the strands inhomogeneous. The inhomogeneities in the DNA molecular chain can also be due to the presence of abasic site-like nonpolar mimic of thymine or external molecules in the sequence or to the fact that the two different base pairs of the real DNA, $A-T$ and $G-C$, combine in different ways constituting the genetic code \cite{lad, cube}. Along the same line, in their study on the dynamics of DNA with periodic and localized inhomogeneities both in stacking and in hydrogen bonds, using the plane base rotator model, Daniel \emph{et al.} \cite{danD, danA} have shown that the inhomogeneity is found to modulate the width and velocity with which the open state configuration travels along the double helical chain. They also demonstrate that the inhomogeneity introduces fluctuations in the open state configuration represented by the kink/antikink-type soliton \cite{danD, danA}. Ag\"uero \emph{et al.} \cite{agu}, using the generalized coherent states approach to averaging the quasi-spin Hamiltonian for DNA, have found that, in a weakly saturating approximation, the formation of compacton/anticompacton pairs for the hydrogen bond displacements is strongly influenced by the inhomogeneity due to the action of an external agent on a specific site of the DNA \cite{agu}.

The results obtained in the above studies demonstrate that, the interplay between nonlinearity and disorder in the DNA through biological processes is yet clearly far from being understood and rather encourage us to extend the study in another case: the case where the inhomogeneities are supposed to be due to the presence of additional molecules such as drugs, mutants, carcinogens or other in specific sites of DNA sequences along the molecular chain and cause damages or mutations on it. The mutations which are the accidental changes observed in the genetic code contains, occur constantly and are due to the actions of endogenous or external agents such as redox-cycling events involving environmental toxic agents and Fenton reactions mediated by heavy metals. Also, reactive oxygen and nitrogen compounds produced by macrophages and neutrophils at sites of inflammation and infection arising as by-products from oxidative respiration can lead to mutations. The above chemical agents can attack DNA, leading to adducts that impair base pairing or block DNA replication and transcription processes, base loss, or DNA single-strand breaks (SSB). Furthermore, when the DNA-replication apparatus encounters a SSB or certain other lesions, double-strand breaks (DSB) are formed \cite{val,kawa,khan,jack}. Cancer usually results from a series of mutations within a single cell. Often, a faulty, damaged, or missing $p53$ gene are to blame. The $p53$ gene provides proteins that stop mutated cells from dividing. Without this protein, cells divide unchecked and become tumours \cite{bie}. The DNA lesions listed above can be gathered in three groups: the benefit mutations which create genetic diversity and keep population healthy, the silent mutations which have no effect at all (such diseases do not manifest) and the mutations which lead to diseases (base loss, SSB or DSB). Such DNA lesions are extremely toxic and difficult to repair. Then, the most promising directions in biophysics is the study of inhomogeneous nonlinear models of DNA, because this can give new interesting relations between the physical nonlinear properties of DNA and its biological functioning. Studies can lead to the discovery of the new mechanisms of regulation of fundamental biological functions of DNA, what can ''bridge'' the nonlinear physics of DNA and medicine \cite{yaku1}. For instance, breather-impurity interactions and its scattering in the PB model was studied in detail by\emph{ Kyle et al.} \cite{kyl}. They show that the impurity can act as a \emph{catalyst} and generates larger excitations during the fusion of the breather mode due to the nonlinearity in the system and the one due to the mass inhomogeneity.

With this in mind, in this work, we aim to study the nonlinear dynamics of the DNA double helix taking into account site dependence inhomogeneities, by considering the PB model. In this case, the DNA dynamics is governed by a perturbed nonlinear Schr\"{o}dinger (NLS) equation and thus, the problem boils down to solve this equation.

The paper is organized as follows. In \secref{model}, we propose the model Hamiltonian  and derive the discrete equations of motion for the in-phase and out-of-phase motions, respectively. Using the multiple-scale expansion in the semi-discrete approximation, the lattice equation is reduced to the Perturbed NLS equation. In \secref{solution}, the solitonic parameters and the first order soliton solution are obtained through the soliton perturbation technique. Inhomogeneity effects on the velocity, wavenumber, angular frequency, shape and position of the soliton solution are discussed. \secref{conclusion} concludes the paper.

\section{\label{model} Model Hamiltonian of DNA dynamics and equations of motion}

We consider the PB model \cite{PB} for DNA denaturation, where the degrees of freedom $x_n$ and $y_n$ associated to each base pair correspond to the displacements of the bases from their equilibrium positions along the direction of the hydrogen bonds that connect the two bases in a pair. A coupling between the base pairs due to the presence of the phosphate groups along the DNA strands is assumed to be inhomogeneous \cite{gra,davi,orns}, so that the Hamiltonian for the model is given by:

\begin{equation}\label{eq1}
\begin{split}
H =&\sum\limits_n \Bigg\{\frac{1}{2}m(\dot x_n^2 + \dot y_n^2)+\frac{1}{2} Kf_n[(x_n-x_{n-1})^2 +(y_n - y_{n - 1})^2] +
V_n(x_n,y_n)\Bigg\},
\end{split}
\end{equation}
where $m$ and $K$ are the nucleotide mass and the elastic coupling constant in the same strand, respectively. The quantity $f_n$ represents the inhomogeneity site dependent character introduced in the transfer of the stacking energy between $n^{th}$ and $(n\pm1)^{th}$ base pairs. It shows that the stacking energy between neighbouring base pairs is site dependent function. As mentioned above, the inhomogeneity represents the intercalation of the compounds (drugs, mutants or carcinogens) between neighbouring base pairs along the DNA molecular chain without any distortion of the strands. The interactions between two bases in a pair are done by the hydrogen bonds which are modelled by the Morse potential given by:
\begin{equation}\label{eq2}
V_n(x_n,y_n)=D\left[e^{-a(x_n-y_n)} - 1 \right]^2,
\end{equation}
where $D$ is the depth of the Morse potential, $a$ is the width of the well. The values of parameters used to describe the motions of the two strands are those from the dynamical and denaturation properties of DNA. They are \cite{oka,pey}: $m=300$ amu, $K=0.06$ eV/\AA$^{2}$, $D=0.03$ $eV$ and $a=4.5$ \AA$^{-1}$. Our system of units (amu, \AA, eV) defines a time unit ($t.u.$) equal to $1.018\times 10^{-14}$ seconds. The variables $u_n$ and $v_n$ are introduced:
\begin{equation}\label{eq3}
u_n = \frac{x_n+y_n}{\sqrt 2}\qquad \mathrm{and} \qquad v_n =\frac{x_n-y_n}{\sqrt 2}.
\end{equation}
Taking into account Eqs. (\ref{eq2}) and (\ref{eq3}), the Hamiltonian of the system becomes
\begin{equation}\label{eq4}
\begin{split}
H&=\sum\limits_n \Bigg\{\frac{1}{2}m\dot u_n^2 + \frac{1}{2}Kf_n(u_n- u_{n - 1})^2\Bigg\}+\sum\limits_n \Bigg\{\frac{1}{2}m\dot
v_n^2+\frac{1}{2}Kf_n(v_n - v_{n - 1})^2+ D\left(e^{-a\sqrt2v_n}-1\right)^2\Bigg\},
\end{split}
\end{equation}
where $u_n$ and $v_n$  represent the in-phase and the out-of-phase motions, respectively. By using the Hamiltonian of \eqref{eq4}, the equation of motions of the system are given by:
\begin{equation}\label{eq5}
m\ddot u_n=\Big[Kf_n\left(u_{n+1}-u_n\right)+Kf_{n-1}\left(u_{n-1}-u_n\right)\Big],
\end{equation}
\begin{equation}\label{eq6}
\begin{split}
m\ddot v_n=\Big[Kf_n\left(v_{n+1}-v_n\right)+Kf_{n-1}\left(v_{n-1}-v_n\right)\Big]+2\sqrt2aDe^{-a\sqrt2v_n}\left(e^{-a\sqrt2v_n}-1\right).
\end{split}
\end{equation}
The equation of variable $u_n(t)$ describes the linear waves (phonons), while the one of variable $v_n(t)$  describes the nonlinear waves (solitons). Hence, we restrict our attention on the second equation of motion (\eqref{eq6}), in which the small amplitude oscillation of the nucleotides is assumed around the bottom of the Morse potential, allowing the following transformation \cite{pey}
\begin{equation}\label{eq7}
v_n= \varepsilon\psi_n,
\end{equation}
where $\varepsilon$ is a small parameter ($\varepsilon\ll1$).
Replacing $v_n$ defined below into \eqref{eq6} and considering the system slightly inhomogeneous, our investigations will be limited to the analysis of the dynamical behaviour of the stretching motion of each base pairs, represented by the solution of \eqref{eq6}. We assume $f_n=(1+\varepsilon^2g_n)$, where $g_n$ is a site-dependent function, which measures the inhomogeneity in the stacking. $g_n$ is treated perturbatively by assuming that, it contributes little enough to the whole DNA dynamics which is dominated by the first and second terms of the following equation obtained up to the third order of the Morse potential,
\begin{equation}\label{eq8}
\begin{split}
\ddot \psi_n&=K_1(\psi_{n+1}-2\psi_n+\psi_{n-1})-\omega_g^2(\psi_n+\varepsilon\alpha\psi_n^2+\varepsilon^2 \beta\psi_n^3) +\varepsilon^2K_1\Big[g_n\left(\psi_{n+1}-\psi_n\right)+g_{n-1}\left(\psi_{n-1}-\psi_n\right)\Big],\\
\end{split}
\end{equation}
where $K_1 = K/m$, $\omega_g^2  = \frac{{4{a^2}D}}{m}$, $\alpha=-\frac{3a}{\sqrt 2}$ and $\beta  = \frac{7a^2}{3}$.
Equation (\ref{eq8}) is the equation describing the dynamics of the out-of-phase motion of a weakly inhomogeneous DNA model. It can be solved directly using the series expansion unknown function method \cite{zdr} or the exponential rational function method \cite{tala}. In this paper, we are looking for the envelope soliton in the small amplitude approximation as a perturbed plane wave solution. For this purpose, it has been shown that the multiple-scale expansion in the semi-discrete approximation is the most adapted technique \cite{oka2, pey, rem}. This approximation is a perturbation technique in which the amplitude is treated in the continuum limit, while the carrier waves are kept discrete. The technique allows the study of the modulation of the wave. Thus, the soliton solution is looking for in the form
\begin{equation}\label{eq9}
\begin{split}
\psi _n =& F_{1,n}e^{i\theta _n}+ \varepsilon \left(F_{0,n}+F_{2,n}e^{2i\theta_n} \right)+C.C.,
\end{split}
\end{equation}
where C.C. stands for complex conjugate and $\theta_n = qnr-\omega t$. The quantities $\omega$, $r$ and $q$ represent the optical frequency of the linear approximation of the base pair vibrations, the distance of neighbouring bases in the same strand, and the wavenumber, respectively. Nonlinear terms in \eqref{eq8} incite one to predict that, through frequency superpositions, the first harmonics of the wave will contain terms in $e^{\pm2i\theta_{n}}$ as well as terms without any exponential dependence. The amplitudes $F_{1,n}$, $F_{0,n}$ and $F_{2,n}$ will be equally considered to change slowly in space and time. For this purpose, the continuum limit approximation and the multiple-scale expansion will be applied on those amplitudes. Thus, the amplitudes are treated as functions of variables according to the new space and time scales $z_i=\varepsilon^iz$ and $T_i=\varepsilon^it$, respectively. Hence, the solution $v_n(t)\rightarrow v(z,t)$, which depends on these new sets of variables is found as a perturbation series of functions. We will consider here that $v(z,t)=\sum\limits_{i=1}^\infty \varepsilon^i\psi_i(z_0, z_1, z_2,..., T_0, T_1, t_2)$. By Taylor expansion, and up to the second order in $\varepsilon$, we obtain for spatial derivatives
\begin{equation}\label{eq10}
\begin{split}
F_{1, n\pm 1} =& F_1,\pm(\varepsilon r)\frac{\partial F_1}{\partial z_1}\pm(\varepsilon r)^2\frac{\partial F_1}{\partial z_2}+
\frac{(\varepsilon r)^2}{2}\frac{\partial^2F_1}{\partial z_1^2} +O((\varepsilon r)^3),
\end{split}
\end{equation}
and their temporal derivatives
\begin{equation}\label{eq11}
\frac{\partial F_{1,n}}{\partial t}=\varepsilon\frac{\partial F_1}{\partial T_1} + \varepsilon^2\frac{\partial F_1}{\partial T_2}+ O(\varepsilon^3).
\end{equation}
The same procedure is used for $F_{0,n}$ and $F_{2,n}$. Using Eqs. (\ref{eq9}), (\ref{eq10}) and (\ref{eq11}) together with \eqref{eq8} and collecting the coefficients for the different powers of ($\varepsilon$, $e^{i\theta_n}$), one obtains the angular frequency and the group velocity of the wave
\begin{equation}\label{eq12}
\begin{split}
\omega^2=\omega_g^2+4K_1\sin^2(qr/2),
\end{split}
\end{equation}
and
\begin{equation}\label{eq13}
\begin{split}
&v_g=\frac{K_1r\sin(qr)}{\omega}.
\end{split}
\end{equation}
The functions $F_0$ and $F_2$ in \eqref{eq9} can be expressed through $F_1$ as
\begin{equation}\label{eq14}
F_0 =-2\alpha|F_1|^2\qquad F_{2}=bF_1^2,
\end{equation}
with
\begin{equation}\label{eq15}
\begin{split}
&b=\frac{\omega_g^2\alpha}{4\omega^2-\varsigma},
\end{split}
\end{equation}
where $\varsigma=\omega_g^2+4K_1\sin^2(qr)$, while $F_1$ is a solution of the following Perturbed NLS equation
\begin{equation}\label{eq16}
\begin{split}
\frac{\partial ^2F_1}{\partial T_1^2} - 2i\omega\frac{\partial F_1}{\partial T_2}=&2i\frac{\partial F_1}{\partial z_2}
K_1r\sin(qr)+K_1r^2\frac{\partial^2F_1}{\partial z_1^2}\cos(qr)-\omega_g^2\alpha(2F_0F_1+2F_1^{*}F_2)\\
&-3\omega_g^2\beta|F_1|^2F_1-4K_1\sin^2(\frac{qr}{2})g(z)F_1.
\end{split}
\end{equation}
After changing to the frame, moving at the group velocity of the carrier wave $v_g$, by defining $\tau=\varepsilon^2t$ and
$X=\varepsilon(z-v_gt)$, we get:
\begin{equation}\label{eq17}
\begin{split}
&i\frac{\partial F_1}{\partial\tau}+P\frac{\partial ^2F_1}{\partial X^2}+2Q|F_1|^2F_1=\frac{2K_1 \sin^2(qr/2)}{\omega}g(X+v_g\tau)F_1,
\end{split}
\end{equation}
where
\begin{equation}\label{eq18}
\begin{split}
&P=\frac{1}{2\omega}[K_1r^2\cos(qr)- v_g^2],\qquad Q=-\frac{\omega_g^2}{4\omega}\Big[2\alpha(-2\alpha+b)+3\beta\Big].
\end{split}
\end{equation}
In the following, we rescale the variable $\tau$ as $\tau$ $\rightarrow$ $(\tau/P)$ and define a new function $G$ such as $F_1(X,\tau) =\sqrt{(P/Q)}$ $G(X,\tau)$. Inserting the above considerations in \eqref{eq17}, we get the Perturbed NLS equation in the form
\begin{equation}\label{eq19}
\begin{split}
i\frac{\partial G}{\partial\tau}+\frac{\partial ^2G}{\partial X^2}+2|G|^2G=\nu Ag(X+v_g\tau)G,
\end{split}
\end{equation}
where $\nu =\sin^2(qr/2)\ll1$ is assumed to be the perturbation parameter, and $A=\frac{2K_1}{P\omega}$, the amplitude of the inhomogeneity.

It should be noted that \eqref{eq19} is the Perturbed NLS equation, regulating the dynamics of the envelope soliton which represents the out-of-phase motion of DNA in the presence of inhomogeneity. In the case of a system without inhomogeneity in the lattice ($g=0$) or when the wave vibrates at the frequency $\omega$ close to $\omega_g$ ($\omega\approx\omega_g$), the above equation is reduced to the following NLS equation
%Schr\"{o}dinger
\begin{equation}\label{eq20}
\begin{split}
i\frac{\partial G}{\partial\tau}+\frac{\partial ^2G}{\partial X^2}+2|G|^2G=0,
\end{split}
\end{equation}
with the well known soliton solution \cite{jia,danE,karp,karp2}
\begin{equation}\label{eq21}
\begin{split}
G(X,\tau)=&2\eta\sech\Big[2\eta\Big((X-X_0)-4c\tau\Big)\Big] e^{i[2c(X-X_0)+4(c^2-\eta^2)\tau+\delta_0]},
\end{split}
\end{equation}
where $\eta$, $c$, $X_0$, $\delta_0$ are four real parameters which represent the height (as well as the width), the velocity, the initial position and initial phase of the propagating soliton, respectively.
\section{\label{solution} Effect of site dependence inhomogeneity on the open state}

In this section, the soliton perturbation technique is used to construct the first-order perturbed soliton solution of \eqref{eq19} and to find the modifications on its parameters due to the inhomogeneity. The technique allows to consider the inhomogeneity as a perturbation term due to the fact that, $\nu$ is a small positive constant measuring the weakness of the perturbation.

Following the work done by \cite{jia, danE, karp,karp2}, \eqref{eq19} is linearized  by transforming the independent variable $\tau$ into
several variables
\begin{equation}\label{eq22}
\begin{split}
&t_n=\nu^n\tau, \qquad \partial_{\tau}=\partial_{t_0}+\nu\partial_{t_1}+\nu^2\partial_{t_2}+...,
\end{split}
\end{equation}
where the subscripts stand for partial differentiation with respect to the time $t_n$.

It is more convenient to represent everything in the coordinate system moving with the soliton. Then, we use $Z$ as a new space independent variable in place of $X$. Under this assumption of quasi-stationary and due to the new time scale introduced below, the soliton parameters $\eta$, $c$, $\xi$, and $\delta$ are now supposed to be function of the slow time variables $t_1$. $c$ and $\eta$ are independent of $t_0$ \cite{jia, danE, karp,karp2}. The one soliton (see \eqref{eq21}) solution of the NLS equation is rewritten here for convenience in the form:
\begin{equation}\label{eq23}
\begin{split}
&G(Z,\tau, \nu)= 2\eta\sech\varphi e^{i\vartheta},
\end{split}
\end{equation}
with
\begin{equation}\label{eq24}
\begin{split}
&\vartheta=\frac{c}{\eta}\varphi-(\delta-\delta_0), \qquad \varphi=2\eta(Z-Z_0)\qquad Z=X-\xi, \qquad \xi_{t_0}=-4c, \qquad
\delta_{t_0}=4(c^2+\eta^2).
\end{split}
\end{equation}
By assuming that the wave propagates at the velocity $c$ close to the quarter of the group velocity, \eqref{eq24} in \eqref{eq19} leads,
\begin{equation}\label{eq25}
\begin{split}
i\frac{\partial G}{\partial\tau}+\frac{\partial ^2G}{\partial Z^2}+2|G|^2G=iR[G],
\end{split}
\end{equation}
where $R[G]=-i\nu Ag(Z)G$.

We assume the solution $G(Z,t_1,\nu)$ to be on the form:
\begin{equation}\label{eq26}
\begin{split}
G(Z,t_1,\nu)=G_0(Z,\tau,\nu)e^{i\vartheta},
\end{split}
\end{equation}
where $G_0(Z,\tau, \nu)$ is the amplitude of the envelope soliton. Hence, \eqref{eq26} in \eqref{eq25} gives the linearized Perturbed NLS equation in the form:
\begin{equation}\label{eq27}
\begin{split}
-4\eta^2G_0+\frac{\partial^2G_0}{\partial Z^2}+2|G_0|^2G_0=\nu F[G_0],
\end{split}
\end{equation}
with
\begin{equation}\label{eq28}
\begin{split}
F[G_0]&=\Big[g(Z)-2\eta (Z-Z_0)c_{t_1}+\Big(2c\frac{\partial Z_0}{\partial t_1}-\frac{\partial \delta_0}{\partial t_1}\Big)\Big]G_0 -i\frac{\partial G_0}{\partial t_1}.
\end{split}
\end{equation}
We use the Poincar\'e expansion type
\begin{equation}\label{eq29}
\begin{split}
&G_0=G_0^{(0)}+\nu G_0^{(1)}+\nu^2G_0^{(2)}+...,\\
&F[G]=F\big[G_0^{(0)}\big]+\nu F\big[G_0^{(0)}, G_0^{(1)}\big]+...
\end{split}
\end{equation}
Inserting \eqref{eq29} into \eqref{eq27} and equating the coefficients of each power of $\nu$, we obtain the following equations at different orders of $\nu$:\\
At the order $\nu^0$, we get the following equation known as stationary NLS equation
\begin{equation}\label{eq30}
\begin{split}
-4\eta^2G_0^{(0)}+\frac{\partial^2G_0^{(0)}}{\partial Z^2}+2|G_0^{(0)}|^2G_0^{(0)}=0,
\end{split}
\end{equation}
where $G_0^{(0)}=2\eta\sech\varphi$;
\\
At the order $\nu^1$, we get the stationary Perturbed NLS equation in the form:
\begin{equation}\label{eq31}
\begin{split}
-4\eta^2G_0^{(1)}+\frac{\partial^2G_0^{(1)}}{\partial Z^2}+2|G^{(0)}|^2\Big(2G^{(1)}+G^{(1)*}\Big)= F[G_0^{(0)}],
\end{split}
\end{equation}
where $^*$ means the complex conjugate.

The term $G^{(1)}_0$ given in \eqref{eq29} is assumed to be on the form:
\begin{equation}\label{eq32}
\begin{split}
G_0^{(1)}(Z,t_1,\nu)=A_1(Z,t_1,\nu)+iB_1(Z,t_1,\nu).
\end{split}
\end{equation}
Introducing \eqref{eq32} into \eqref{eq31} gives:
\begin{equation}\label{eq33}
\begin{split}
&\hat{L}_1A_1\equiv-4\eta^2A_1+\frac{\partial^2A_1}{\partial Z^2}+6|G_0^{(0)}|^2A_1=ReF\big[G_0^{(0)}\big],\\
&\hat{L}_2B_1\equiv-4\eta^2B_1+\frac{\partial^2B_1}{\partial Z^2}+2|G_0^{(0)}|^2B_1=ImF\big[G_0^{(0)}\big],
\end{split}
\end{equation}
with
\begin{equation}\label{eq34}
\begin{split}
&ReF\big[G_0^{(0)}\big]=\Big[g(Z)-2\eta (Z-Z_0)c_{t_1}+\Big(2c\frac{\partial Z_0}{\partial t_1}-\frac{\partial \delta_0}{\partial t_1}\Big)\Big]G_0^{(0)}, \qquad ImF\big[G_0^{(0)}\big]=-\frac{\partial G_0^{(0)}}{\partial t_1},
\end{split}
\end{equation}
and
\begin{equation}\label{eq35}
\begin{split}
&\hat{L}_1=-4\eta^2+\frac{\partial^2}{\partial Z^2}+6|G_0^{(0)}|,\qquad \hat{L}_2=-4\eta^2+\frac{\partial^2}{\partial Z^2}+2\big|G_0^{(0)}\big|,
\end{split}
\end{equation}
$\hat{L}_1$ and $\hat{L}_2$ are two self-adjoint operators.

\subsection{Variation of the solitonic parameters}

In this part, the variation of the parameters of the soliton is evaluated by assuming that, the soliton propagates with an amplitude
(as well as a width) $\eta=\eta_0$ and a velocity $c=c_0$, when the perturbation is switched off. To evaluate the modifications of the solitonic parameters, it is necessary to solve the homogeneous parts of \eqref{eq33} which admit $\phi_1$ and $\psi_2$ as solutions, respectively. The non-secularity conditions \cite{jia,karp,karp2} give us the following four important formulas which determine how the solitonic parameters (i.e. the amplitude, the position, the velocity and the angular frequency) are modified by the inhomogeneity
\begin{equation}\label{eq36}
\begin{split}
\eta_{t_1}=\frac{1}{2}\int_{-\infty}^{+\infty}Re\Big[R\big[G_0^{(0)}\big]\Big]\phi_1(\varphi)d\varphi,
\end{split}
\end{equation}
\begin{equation}\label{eq37}
\begin{split}
\xi_{t_1}=\frac{1}{4\eta^2}\int_{-\infty}^{+\infty}Re\Big[R\big[G_0^{(0)}\big]\Big]\phi_2(\varphi)d\varphi,
\end{split}
\end{equation}
\begin{equation}\label{eq38}
\begin{split}
c_{t_1}=-\frac{1}{2}\int_{-\infty}^{+\infty}Im\Big[R\big[G_0^{(0)}\big]\Big]\psi_2(\varphi)d\varphi,
\end{split}
\end{equation}
\begin{equation}\label{eq39}
\begin{split}
\delta_{t_1}=2c\xi_{t_1}-\frac{1}{2\eta}\int_{-\infty}^{+\infty}Im\Big[R\big[G_0^{(0)}\big]\Big]\psi_1(\varphi)d\varphi,
\end{split}
\end{equation}
with
\begin{equation}\label{eq40}
\begin{split}
&\phi_1(\varphi)=\sech\varphi, \qquad \phi_2(\varphi)=\varphi\sech\varphi,\qquad \psi_1(\varphi)=(1-\varphi\tanh\varphi)\sech\varphi, \qquad
\psi_2(\varphi)=\tanh\varphi\sech\varphi,
\end{split}
\end{equation}
where $\varphi=2\eta(Z-Z_0)$. $Re\Big[R[G_0^{(0)}]\Big]$ and $Im\Big[R[G_0^{(0)}]\Big]$ are the real and imaginary parts of $R[G_0^{(0)}]$ given in \eqref{eq25}.

To find the variation of the soliton parameters explicitly, we have to evaluate the integrals found in the right-hand sides of Eqs. [(\ref{eq36})-(\ref{eq39})], which can be carried out only on supplying the specific forms of $f(\varphi)$. Hence, we consider the localized inhomogeneity in the form of the hyperbolic tangent function and the exponential inhomogeneity in the form of exponential function separately. The results give the time dependence of the parameters of the soliton as:
\begin{equation}\label{eq41}
\begin{split}
&\eta_{t_1}=0, \qquad \eta=\eta_0,\\
&\xi_{t_1}=0, \qquad \xi_{\tau}=\xi_{t_0},\\
&c_{t_1}=\eta A\int_{-\infty}^{+\infty}f(\varphi)\sech^2\varphi\tanh\varphi d\varphi,\\
&\delta_{t_1}=A\int_{-\infty}^{+\infty}f(\varphi)\sech^2\varphi(1-\varphi\tanh\varphi)d\varphi.\\
\end{split}
\end{equation}
The first equation of the \eqref{eq41} shows that the amplitude (as well as the width) of the soliton remains constant, showing that the number of base pairs which take part in the opening process remains constant during the propagation. We observe in the second equation that the position of the soliton is not affected by the inhomogeneities. However, from the third and fourth equations, we found that the velocity and the angular frequency of the soliton can get a correction depending of the type of the inhomogeneities in the lattice.

The localized inhomogeneity can represent the intercalation of a compound between neighbouring base pairs or the presence of a defect
on an abasic site in the DNA chain. In order to understand the effects of this type of inhomogeneity in the lattice, we substitute $f(\varphi)=\tanh(\varphi)$, in \eqref{eq41}, and after integration, we obtain $c_{t_1}=\frac{2}{3}\eta A$ and $\delta_{t_1}=0$, which can be written in terms of the original time variable $\tau$ by using the expression $c_\tau=c_{t_0}+\nu c_{t_1}$ and
$\delta_\tau=\delta_{t_0}+\nu \delta_{t_1}$ as:
\begin{equation}\label{eq42}
\begin{split}
&c=c_0+\frac{2}{3}(\nu A)\eta\tau,\qquad \Omega_\nu\equiv\delta_\tau=4(c_0^2+\eta^2)+\frac{8}{3}(\nu A)\eta\Big[2c_0\tau+(\nu A)\eta\tau^2\Big].\\
\end{split}
\end{equation}
Equation (\ref{eq42}) demonstrates that, the localized inhomogeneity in the above form affects the velocity and the angular frequency of the soliton. They get a correction during the crossing of this inhomogeneity. The nature of the correction depends of the nature of the inhomogeneity represented by the sign of its amplitude $A$ which can be either positive or negative. When $A$ is positive ($A>0$), the inhomogeneity corresponds to an energetic barrier and on the other hand, when $A$ is negative ($A<0$), the inhomogeneity behaves as a potential well \cite{danA}.

\begin{description}
  \item When $A$ is greater than zero, the correction is positive. That makes the velocity and the angular frequency increasing during the crossing of the inhomogeneity. The amount of these increments is proportional to the initial height (as well as the initial width) of the soliton. The higher the initial amplitude of the soliton, the greater the increments. The increasing in the velocity of the soliton helps to overcome the barrier due to the inhomogeneity and the soliton will propagate easily along the inhomogeneous DNA chain without formation of a bound state. Reporting these observations in \eqref{eq23}, we notice that during the crossing of the inhomogeneity, the soliton propagates and vibrates quickly, and the DNA breathing mode becomes faster.
  \item In the case of $A$ less than zero, the correction is negative. That makes the frequency and the velocity decreasing. The decreasing of the velocity slows down the soliton. The soliton stops and vanishes when the original time satisfies the condition $\tau=t_0=\frac{-c_0}{B_i(\nu A)\eta}$, where $B_i$ is a constant which depends on the inhomogeneity. The propagating time of the soliton depends on its initial velocity and width. The wider the soliton, the greater the propagating time.
  \item When $A=0$, the velocity and the angular frequency of the soliton remain constant.
\end{description}
In all the above cases, the soliton which is a coherent structure formed by involving a few base pairs, moves along the helical chain in the form of bubble without dissipation or any other form of deformation is found stable.

\begin{figure}[H]
\centering
\includegraphics[width=3in]{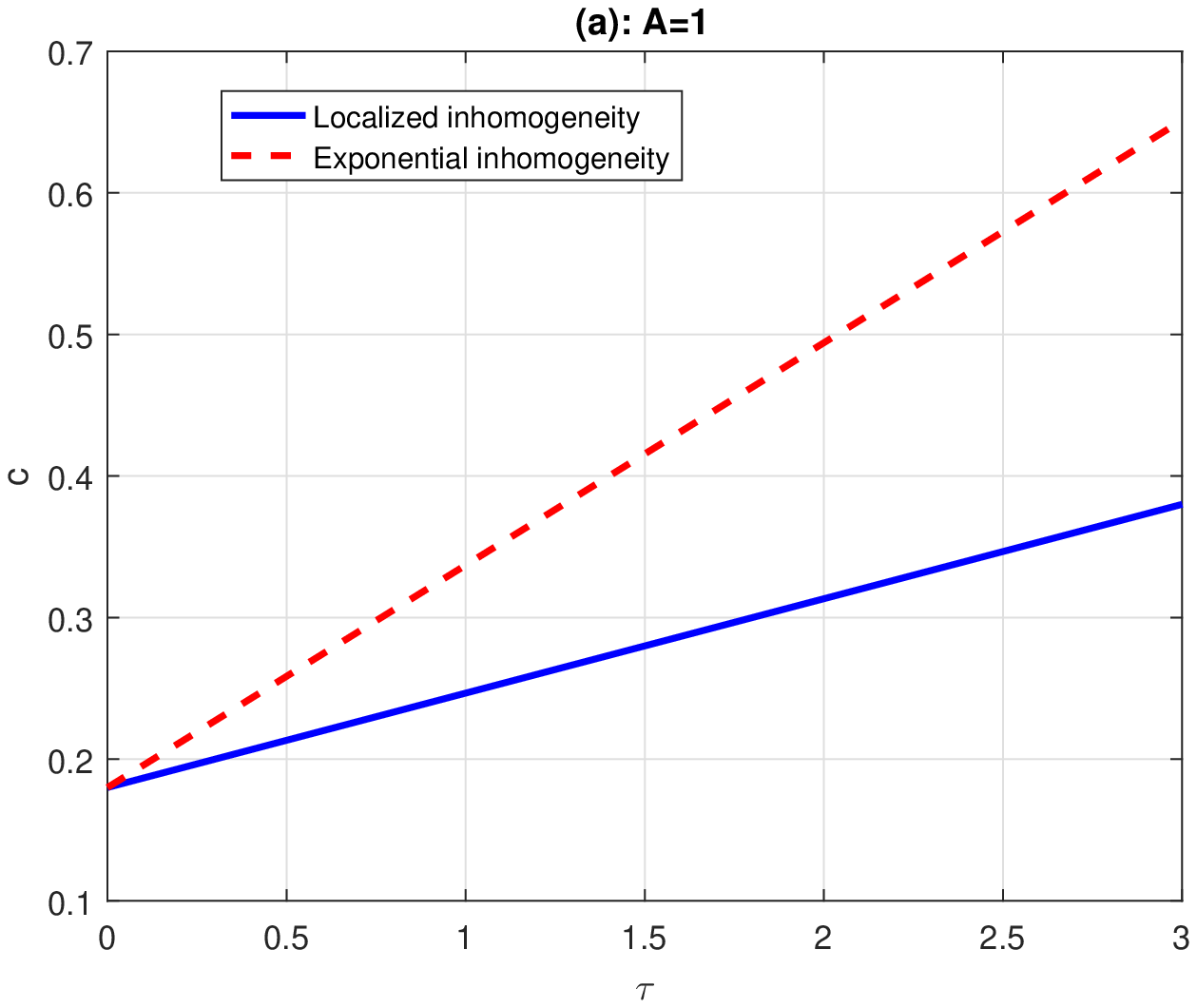}
\includegraphics[width=3in]{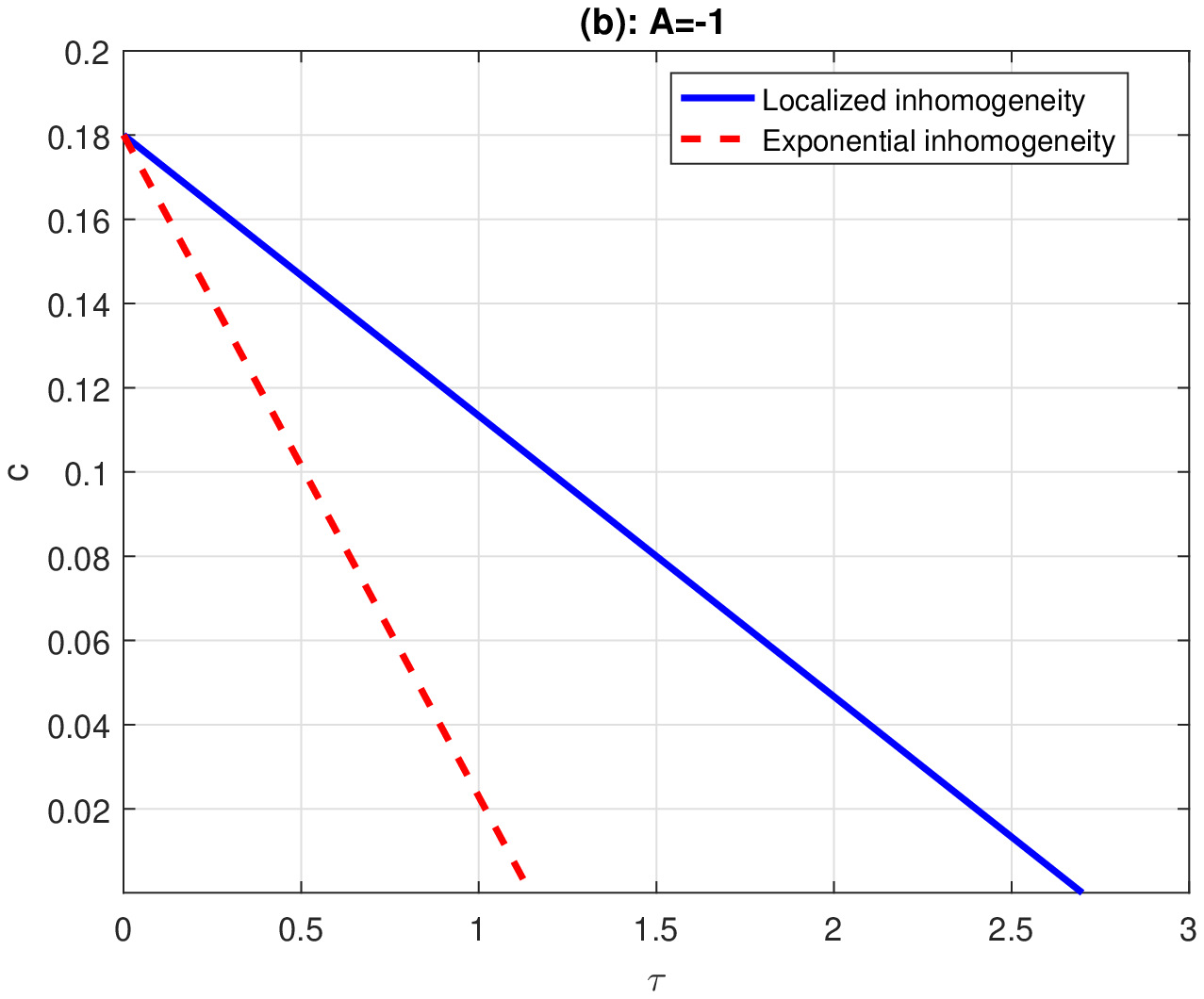}
\caption{Variation of the velocity of the soliton as a function of the time, depending on the nature of the inhomogeneity in the lattice for $\nu=0.1$, $c_0=0.18$, and $\eta=1$.}\label{fig1}
\end{figure}

\begin{figure}[H]
\centering
\includegraphics[width=3in]{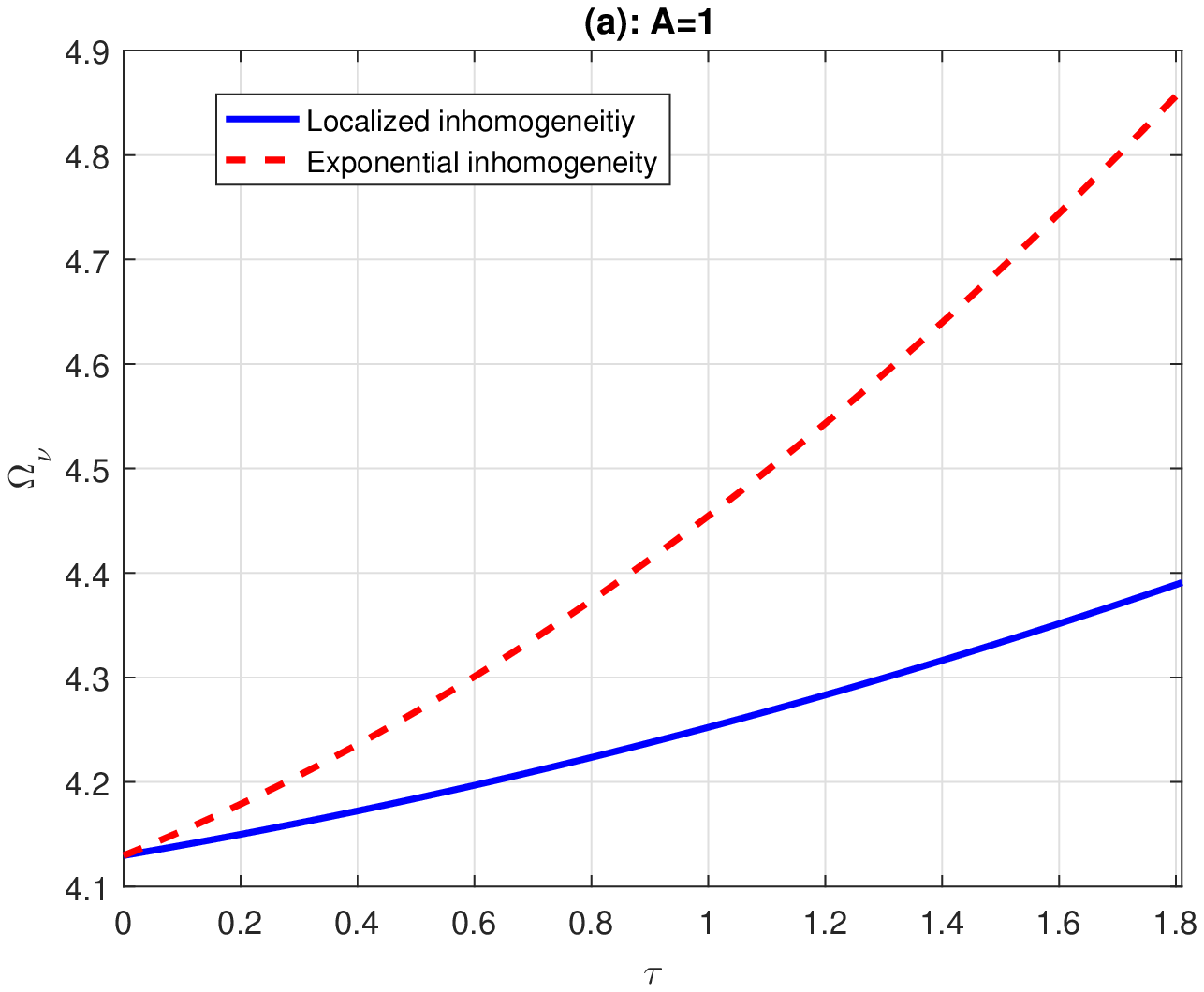}
\includegraphics[width=3in]{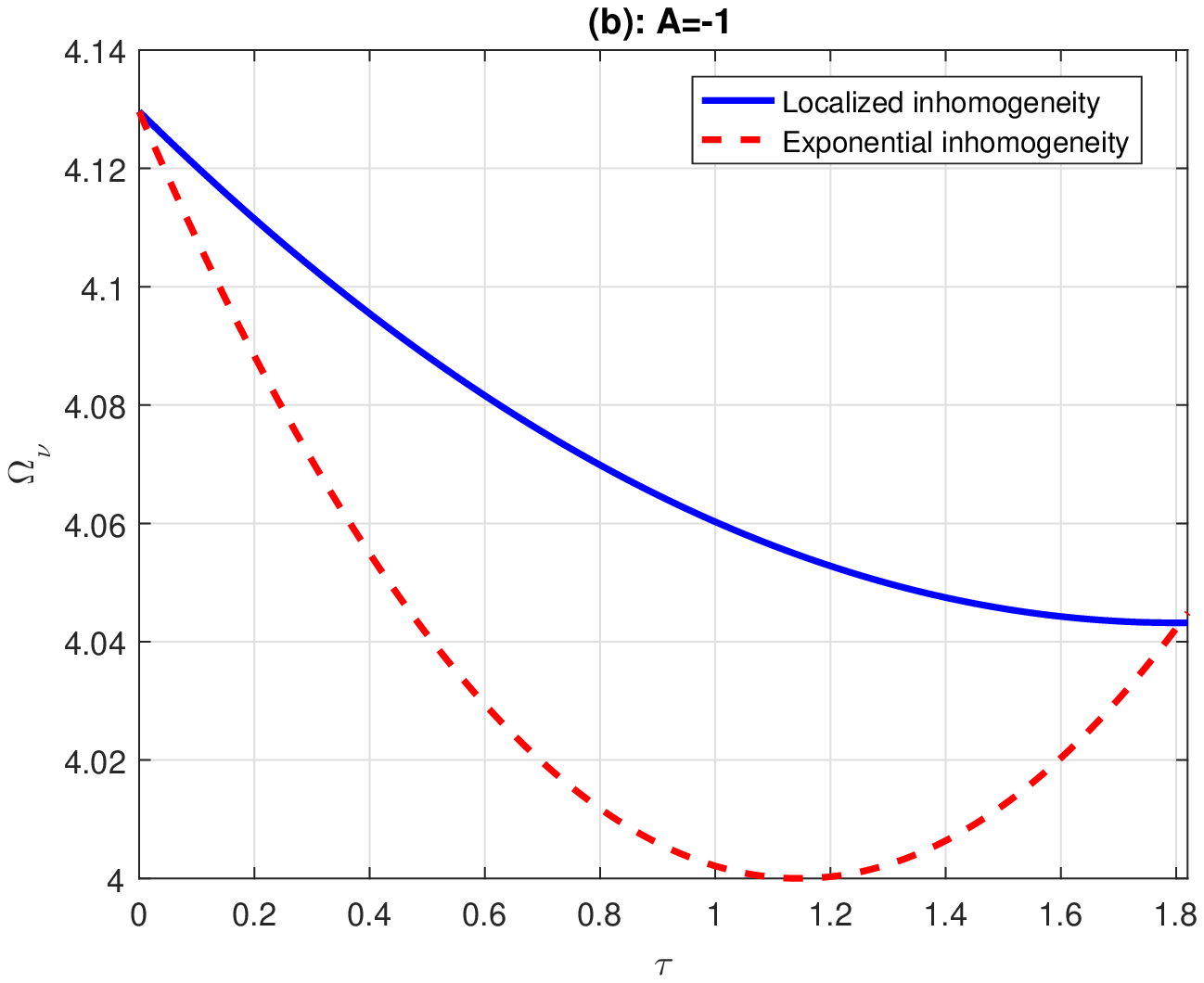}
\caption{Variation of the angular frequency of the soliton as a function of the time, depending on the nature of the inhomogeneity in the lattice. The parameters are the same as in \figref{fig1}.}\label{fig2}
\end{figure}

Next,  we consider $f(\varphi)=e^{\varphi}$. This case corresponds to the physically interesting problem of the exponential distribution of
similar molecules along the DNA helical chain. We have $c_{t_1}=\frac{\pi}{2}\eta A$ and $\delta_{t_1}=0$. Written in the original time variable, we get
\begin{equation}\label{eq43}
\begin{split}
&c=c_0+\frac{\pi}{2}(\nu A)\eta\tau,\qquad
\Omega_\nu\equiv\delta_\tau=4(c_0^2+\eta^2)+2\pi(\nu A)\eta\Big[2c_0\tau+\frac{\pi}{2}(\nu\eta)A\tau^2\Big].
\end{split}
\end{equation}
The same observations as in the previous type of inhomogeneity are also done here. We notice from the results that considering the localized and exponential inhomogeneities, the propagating velocity and frequency of the soliton undergo a correction during the crossing of the inhomogeneities. In Figures (\ref{fig1}) and (\ref{fig2}), the time-evolution of the velocity and angular frequency of the soliton are depicted as functions of the type and nature of the inhomogeneities in the lattice. The figures show that the correction terms can be positive or negative according to the nature of the inhomogeneities.

When comparing \eqref{eq42} and \eqref{eq43}, we notice that, the absolute value of the correction terms is greater in the case of exponential inhomogeneity as can be seen in  the above Figures. This is because in this case, the inhomogeneities occur exponentially in the entire chain of the DNA molecule in terms of external agents.
\subsection{First-order perturbed soliton solution}

In this part of our work, we will seek for the breatherlike soliton solution for \eqref{eq25}. As we know, once the seed solution is chosen as breather soliton solution via \eqref{eq23}, the first-order perturbed breatherlike soliton solution can be constructed using the soliton perturbation technique by solving \eqref{eq33}. Then, we report the solutions $A_1$ and $B_1$ in \eqref{eq32}. We are solving first the homogeneous part of the first equation in \eqref{eq33}, which admits the following two particular solutions:
\begin{equation}\label{eq44}
\begin{split}
A_{11}=\sech\varphi\tanh\varphi,
\end{split}
\end{equation}
and
\begin{equation}\label{eq45}
\begin{split}
A_{12}&=\frac{1}{2\eta}\Big[\frac{3}{2}\varphi \sech\varphi\tanh\varphi+\frac{1}{2}\tanh\varphi\sinh\varphi-\sech\varphi\Big].
\end{split}
\end{equation}
Then, the general solution can be obtained by using the following formula:
\begin{equation}\label{eq46}
\begin{split}
A_{1}&=C_1A_{11}+C_2A_{12}-\frac{1}{2\eta}A_{11}\int_{-\infty}^{\varphi}
A_{12}ReF\big[G_0^{(0)}\big]d\varphi+\frac{1}{2\eta}A_{12}\int_{-\infty}^{\varphi}A_{11}ReF\big[G_0^{(0)}\big]d\varphi,
\end{split}
\end{equation}
where $C_1$ and $C_2$ are two arbitrary constants to be determined. We construct the solution $A_1$, by substituting the expressions of
$A_{11}$, $A_{12}$ and $ReF(G_0^{(0)})$ in \eqref{eq46}. After the evaluation of the integrals, we obtain:
\begin{equation}\label{eq47}
\begin{split}
A_1=&-\frac{1}{2\eta}\Big[C_2-\frac{1}{2}\Big(2c\frac{\partial Z_0}{\partial t_1}-\frac{\partial\delta_0}{\partial t_1}\Big)\Big]\sech\varphi +\Big[C_1+\frac{3C_2}{4\eta}\varphi-\frac{1}{4\eta}\varphi\Big(2c\frac{Z_0}{\partial t_1}-\frac{\partial\delta_0 }{\partial t_1}\Big)\Big]\sech\varphi\tanh\varphi\\
&+\frac{1}{8\eta^2}c_{t_1}(1-\tanh\varphi)\Big[1+\varphi e^{2\varphi} (1-\tanh\varphi)\Big]\Big[3\varphi \sech\varphi\tanh\varphi +\tanh\varphi\sinh\varphi-2\sech\varphi\Big]\\
&+\frac{1}{8\eta^2}c_{t_1}\varphi^2\sech\varphi\tanh\varphi \Big[1-3e^{2\varphi}(1-\tanh\varphi)\Big] +\frac{C_2}{4\eta}\sinh\varphi\tanh\varphi -I+J,
\end{split}
\end{equation}
where $I$ and $J$ are the integrals depending on the form of the inhomogeneities and are given by:
\begin{equation}\label{eq48}
\begin{split}
&I=\frac{1}{2\eta} A A_{11}\int_{-\infty}^{\varphi}A_{12}f(\varphi)G_0^{(0)}d\varphi,\qquad J=\frac{1}{2\eta}A A_{12}\int_{-\infty}^{\varphi}A_{11}f(\varphi)G_0^{(0)}d\varphi\\
\end{split}
\end{equation}
The term in $\frac{C_2}{4\eta}\sinh\varphi \tanh\varphi$ is a secular term which makes the solution unbounded. The above term can be removed if we assume that the arbitrary constant $C_2=0$. Using the boundary conditions $A_1|_{\varphi=0}=\mu$ and $A_{1\varphi}|_{\varphi=0}=0$,
we obtain $\frac{1}{4\eta}\Big(2c\frac{\partial Z_0}{\partial t_1}-\frac{\partial\delta_0}{\partial t_1}\Big) =\mu+\frac{1}{8\eta^2}c_{t_1}$ and $C_1=0$. By using the above results in \eqref{eq47}, we get the general solution in the form:
\begin{equation}\label{eq49}
\begin{split}
A_1=&\Big(\mu+\frac{1}{8\eta^2}c_{t_1}\Big)\Big[1-\varphi\tanh\varphi\Big]\sech\varphi+\frac{1}{8\eta^2}c_{t_1}\varphi^2
\sech\varphi\tanh\varphi\Big[1-3e^{2\varphi}(1-\tanh\varphi)\Big]\\
&+\frac{1}{8\eta^2}c_{t_1}(1-\tanh\varphi)\Big[1+\varphi e^{2\varphi}(1-\tanh\varphi)\Big]\Big[3\varphi\sech\varphi\tanh\varphi
+\tanh\varphi\sinh\varphi-2\sech\varphi\Big]\\
&-I+J.
\end{split}
\end{equation}
Now, we give the solution of the second equation in \eqref{eq33}. Its homogeneous part admits the particular solutions given by:
\begin{equation}\label{eq50}
\begin{split}
B_{11}=\sech\varphi,\quad B_{12}=\frac{1}{4\eta}(\varphi\sech\varphi+\sinh\varphi),
\end{split}
\end{equation}
while the general form of $B_1$ is given by:
\begin{equation}\label{eq51}
\begin{split}
B_{1}&=C_3B_{11}+C_4B_{12}-\frac{1}{2\eta}B_{11}\int_{-\infty}^{\varphi}B_{12}ImF(G_0^{(0)})d\varphi +\frac{1}{2\eta}B_{12}\int_{-\infty}^{\varphi}B_{11}ImF(G_0^{(0)})d\varphi,
\end{split}
\end{equation}
where $C_3$ and $C_4$ are two arbitrary constants to be determined. We follow the same procedure by removing the secular terms ($C_3=C_4=0$). We get also $\frac{\partial Z_0}{\partial t_1}=0$. Applying the following boundary conditions $B_1|_{\varphi=0}=B_{1\varphi}|_{\varphi=0}=0$, we have
\begin{equation}\label{eq52}
\begin{split}
B_{1}=0.
\end{split}
\end{equation}
Taking into account the above results, the general form of the envelope soliton is
\begin{equation}\label{eq53}
\begin{split}
G_0(\varphi,\tau,\nu)=&2\eta\sech\varphi+\nu\Bigg\{\Big(\mu+\frac{1}{8\eta^2}c_{t_1}\Big)\Big[1-\varphi\tanh\varphi\Big]\sech\varphi +\frac{1}{8\eta^2}c_{t_1}\varphi^2\sech\varphi\tanh\varphi\Big[1-3e^{2\varphi}(1-\tanh\varphi)\Big]\\
&+\frac{1}{8\eta^2}c_{t_1}(1-\tanh\varphi)\Big[1+\varphi e^{2\varphi}(1-\tanh\varphi)\Big]\Big[3\varphi\sech\varphi\tanh\varphi +\tanh\varphi\sinh\varphi-2\sech\varphi\Big]-I+J\Bigg\},
\end{split}
\end{equation}
where $I$ and $J$ depend on the type of inhomogeneity and are given in \eqref{eq48}. The effects of the inhomogeneities on the shape of the
first order perturbed soliton is therefore reduced to the integration of $I$ and $J$.

We construct the first order perturbed soliton solution by substituting the corresponding values of $c_{t_1}$, $\delta{t_1}$ for each expression of $f(\varphi)$ given below, and evaluate the integrals $I$ and $J$, which involve very lengthy algebra. Thus, in the case $f(\varphi)=\tanh(\varphi)$, we use the values of  $c_{t_1}$ and $\delta_{t_1}$ of \eqref{eq42}. \eqref{eq53} becomes,
\begin{equation}\label{eq54}
\begin{split}
G_0=&2\eta\sech\varphi+\nu\Bigg\{\Big(\mu+\frac{\eta A}{12\eta^2}\Big)(1-\varphi\tanh\varphi)\sech\varphi
+\frac{\eta A}{12\eta^2}\varphi^2\sech\varphi\tanh\varphi\Big[1-3e^{2\varphi}(1-\tanh\varphi)\Big]\\
&+\frac{\eta A}{12\eta^2}(1-\tanh\varphi)\Big[1+\varphi e^{2\varphi} (1-\tanh\varphi)\Big]\Big[3\varphi\sech\varphi\tanh\varphi +\tanh\varphi\sinh\varphi-2\sech\varphi\Big]\\
&-\frac{1}{2\eta}A\sech\varphi\tanh\varphi\Big[\varphi-\frac{1}{8}(1-\tanh\varphi)^3(\varphi+e^{2\varphi}+3\varphi e^{4\varphi} +e^{4\varphi})-\frac{1}{2}\ln(e^{2\varphi}+1)-\frac{1}{4}\tanh^2\varphi\\&+\frac{1}{2}\sech^2\varphi\Big]+\frac{1}{6\eta}
A(1-\sech^2\varphi)\Big[\frac{3}{2}\varphi\sech\varphi\tanh^2\varphi+\frac{1}{2}\tanh^2\varphi\sinh\varphi-\sech\varphi\tanh\varphi\Big]\Bigg\}.
\end{split}
\end{equation}
We then repeat the same procedure for constructing the first order perturbed soliton solution in the case of $f(\varphi)=e^\varphi$ and we obtain the perturbed soliton solution as:
\begin{equation}\label{eq55}
\begin{split}
G_0=&2\eta\sech\varphi+\nu\Bigg\{\Big(\mu+\frac{\pi\eta A}{16\eta^2}\Big)(1-\varphi\tanh\varphi)\sech\varphi
+\frac{\pi\eta A}{16\eta^2}\varphi^2\sech\varphi\tanh\varphi\Big[1-3e^{2\varphi}(1-\tanh\varphi)\Big]\\
&+\frac{\pi\eta A}{16\eta^2}(1-\tanh\varphi)\Big[1+\varphi e^{2\varphi}(1-\tanh\varphi)\Big]
\Big[3\varphi \sech\varphi\tanh\varphi+\tanh\varphi\sinh\varphi-2\sech\varphi\Big]\\
&+\frac{1}{4\eta}A\sech\varphi\tanh\varphi\Big[\frac{3}{4}e^{\varphi}(1-\tanh\varphi)^2
(\varphi+3\varphi e^{2\varphi}+e^{2\varphi}+1)+3\tanh\sinh\varphi-4\cosh\varphi-\sinh\varphi\Big]\\
&-\frac{1}{4\eta}A\Big(3\varphi\sech\varphi\tanh\varphi+\tanh\varphi\sinh\varphi -2\sech\varphi\Big)\Big(\frac{1}{2}\tanh\varphi\sech\varphi-\tanh\varphi\sinh\varphi+\cosh\varphi-\arctan(e^{\varphi})\Big)\Bigg\}.
\end{split}
\end{equation}
Equations (\ref{eq54}) and (\ref{eq55}) are the breatherlike solitons which represent the envelope of the soliton solutions. They describe the open state configurations in the individual strand of the DNA, which collectively represents bubbles moving along the inhomogeneous DNA molecule.

\begin{figure}[H]
\centering
\includegraphics[width=3in]{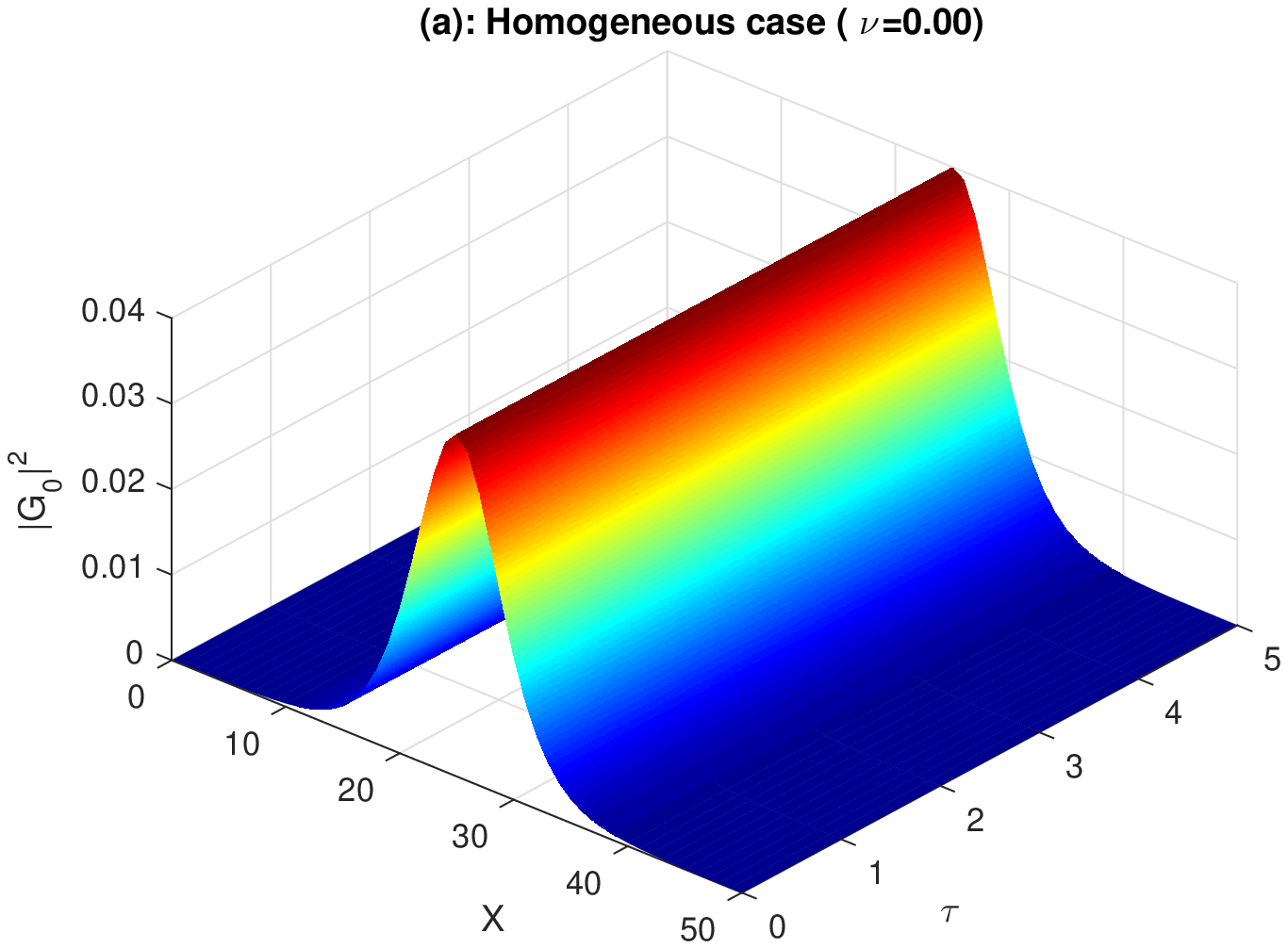}
\includegraphics[width=3in]{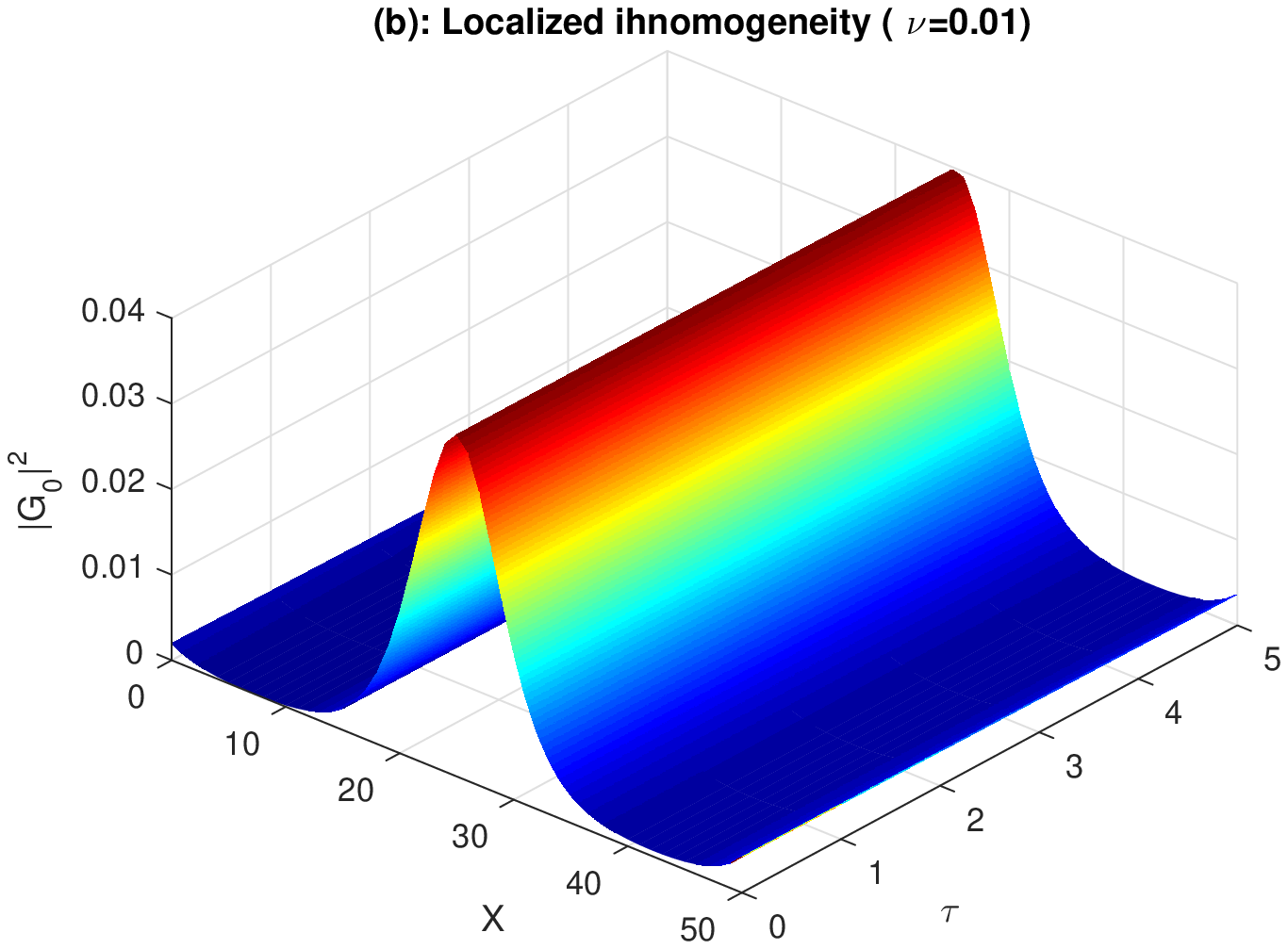}
\includegraphics[width=3in]{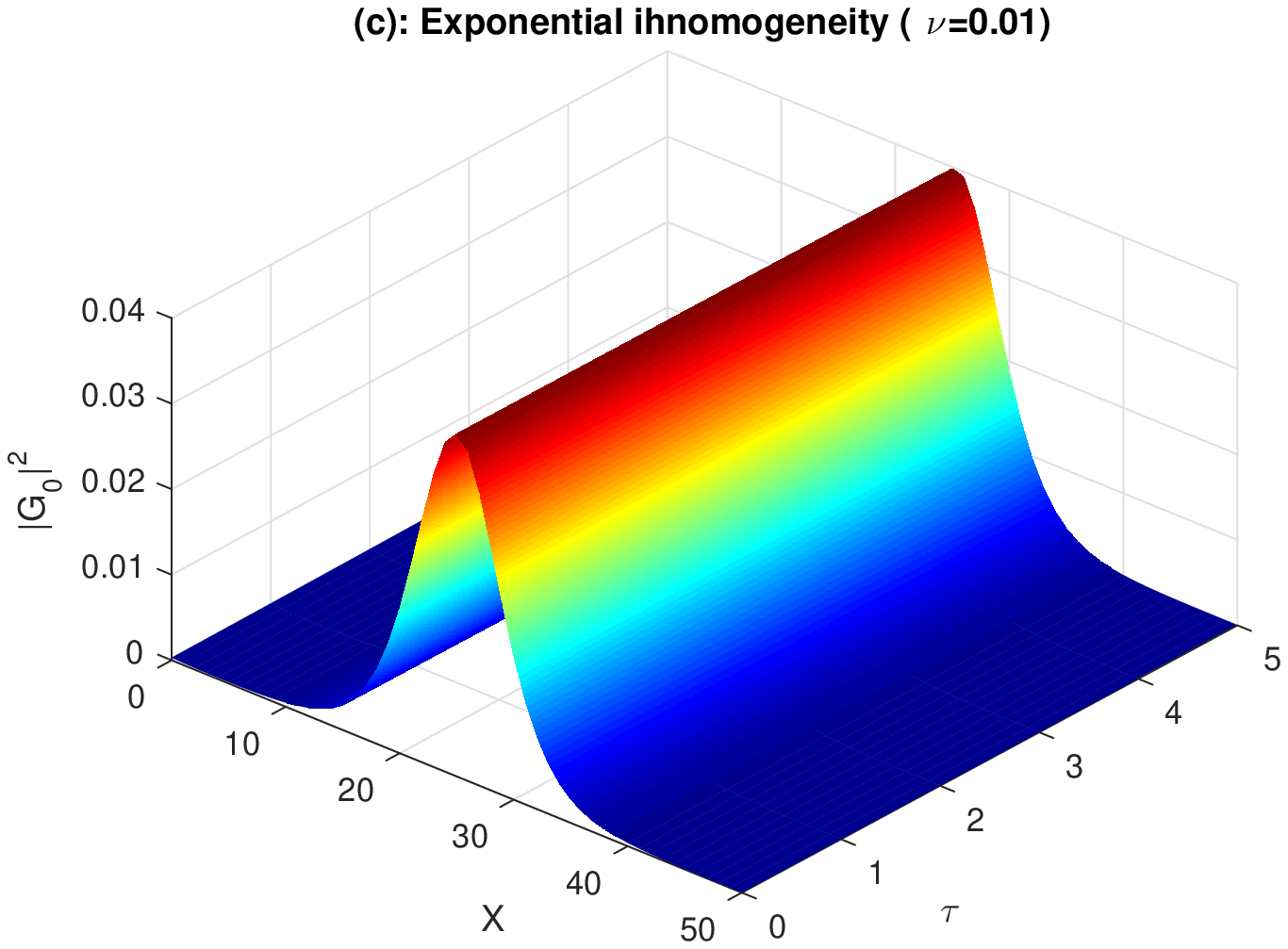}
\caption{Dynamical evolutions of the first-order perturbed breatherlike solitons [see (\ref{eq54})-(\ref{eq55})], with the parameters: $\varepsilon=0.5$, $c_0=0.01$, $\eta=0.1$, $\nu=0.01$, $\mu=0$, $Z_0=0$ and $A=1$.}\label{fig3}
\end{figure}

We have depicted in \figref{fig3}, the $3D$ schematic representation of the square of the envelope solitons as functions of the type of inhomogeneity in the lattice. \figref{fig3}a presents the breather soliton moving in the homogeneous DNA chain. The case of localized inhomogeneity is depicted in \figref{fig3}b, while the case of exponential inhomogeneity is depicted in \figref{fig3}c. We notice that, in both inhomogeneous cases, the robust nature of the breatherlike soliton is not modified as it propagates along the DNA molecule. The global shape of the molecule with inhomogeneities is merged with its shape without taking into account the inhomogeneities. Similar results were observed by Mvogo \emph{et al.} \textbf{\cite{mvo}} in their investigations on the effects of localized inhomogeneity in the form of hyperbolic secant function in the $\alpha$-helical proteins chain. Also we found in Ref. \cite{saha} that, using the numerical methods, the authors demonstrate that neither the opening of base pairs in DNA molecule nor the topological character of the breatherlike soliton are affected by the localized inhomogeneity in the form of hyperbolic secant function introduced in stacking and hydrogen bonding energies of DNA.

Considering \eqref{eq7} together with \eqref{eq9}, we assume that the configuration of the molecule can be fully described by the following soliton solution with a rescaled time
\begin{equation}\label{eq56}
\begin{split}
v_n=&\varepsilon G_0\cos(Rnr-\Omega t),
%v1=2.*epsilon*eta.*sech(2.*epsilon*eta.*((n-(N/2)).*r-Ve1.*t))
\end{split}
\end{equation}
where
\begin{equation}\label{eq57}
\begin{split}
R=q+2\varepsilon c,\quad \Omega=\omega+2\varepsilon cv_g+\Omega_\nu.
%Ve1=vg+4*epsilon*c1
\end{split}
\end{equation}
$R$ and $\Omega$ are the \textquotedblleft effective \textquotedblright wavenumber and angular frequency of the soliton solution, respectively. $G_0$, $c$ and $\Omega_\nu$ represent the envelope soliton, its velocity and angular frequency, respectively. They depend on the type of inhomogeneity and are given by [\eqref{eq42} and \eqref{eq54}] and [\eqref{eq43} and \eqref{eq55}] for the localized and the exponential inhomogeneities, respectively. \eqref{eq56} represents the stretching of the base pairs also known as the breathing modes experimentally observed in DNA molecule \cite{putn,proh}. Since the wave velocity $c$ depends on the type of inhomogeneity in the lattice, from the first equation of \eqref{eq57}, we notice that the wavenumber $R$ is a function of the inhomogeneities. It changes as time progressed due to the inhomogeneities.

\begin{figure}[H]
\centering
\includegraphics[width=3in]{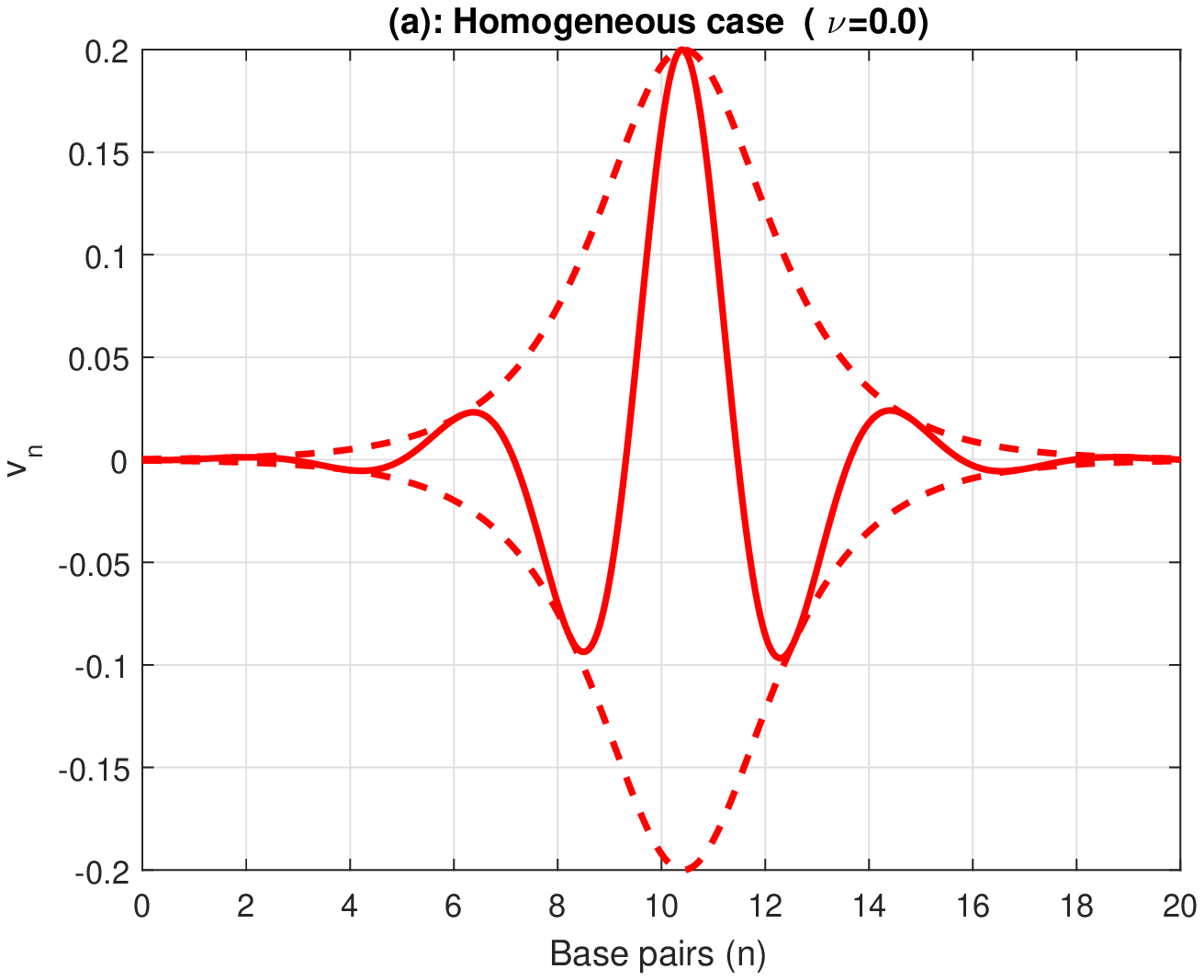}
\includegraphics[width=3in]{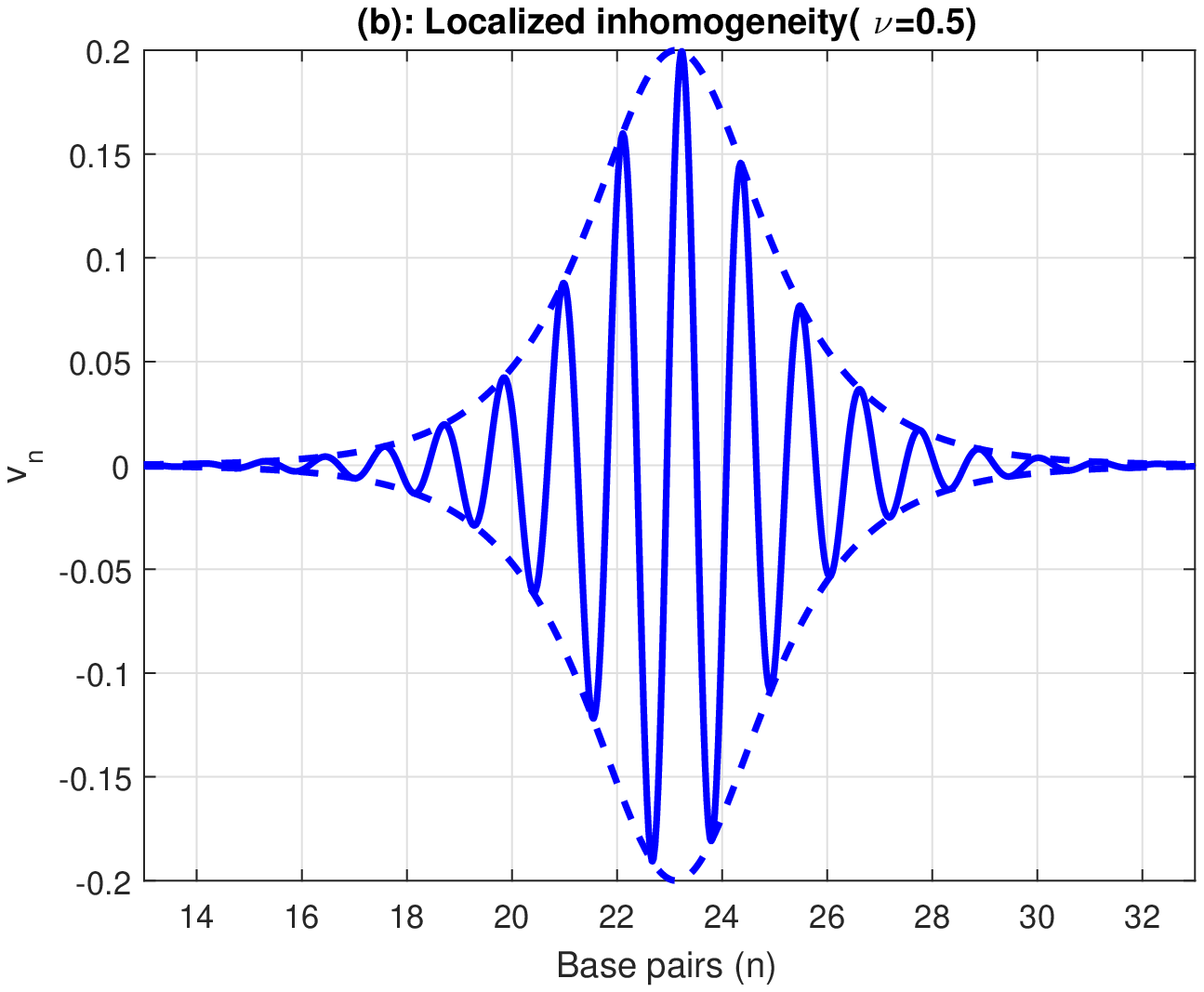}
\includegraphics[width=3in]{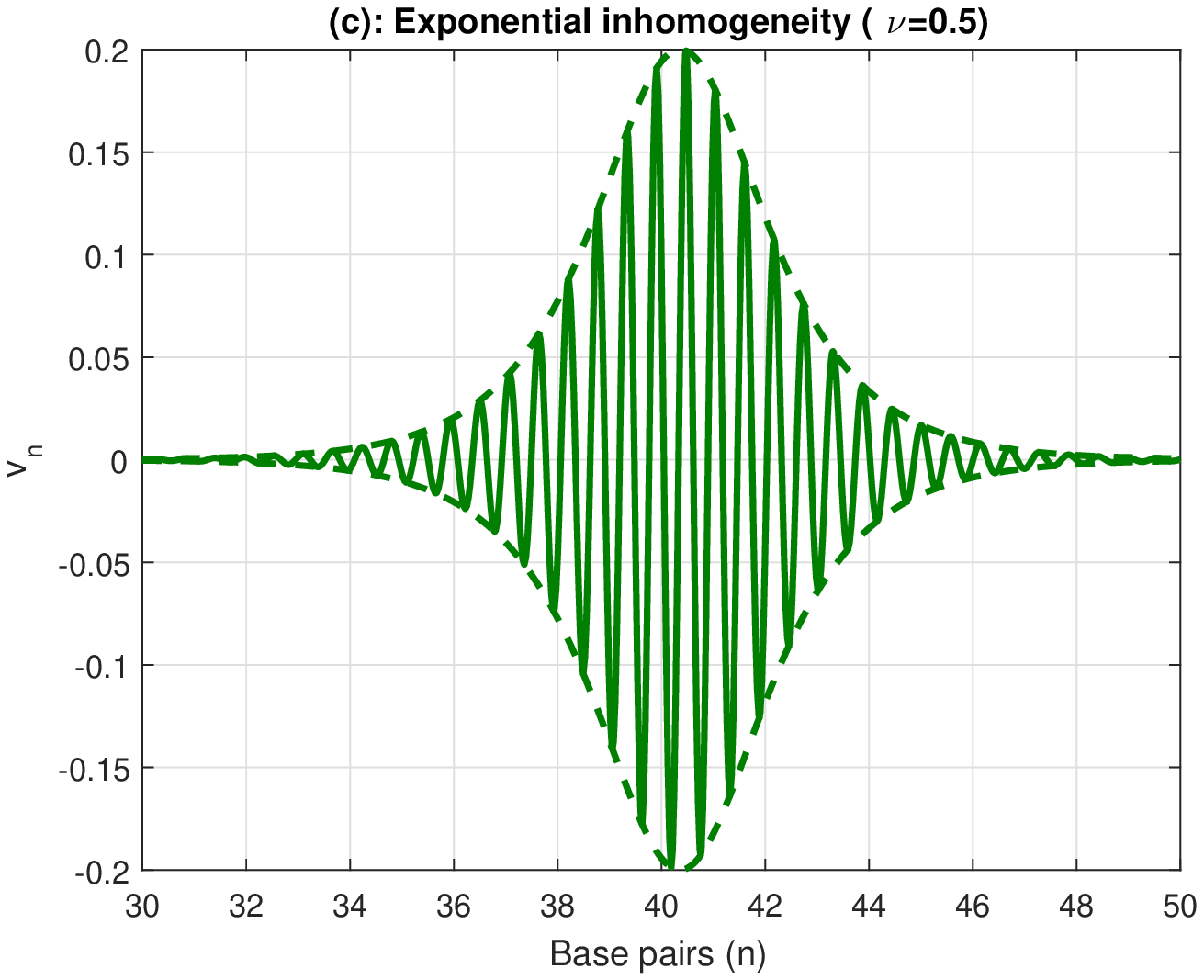}
\includegraphics[width=3in]{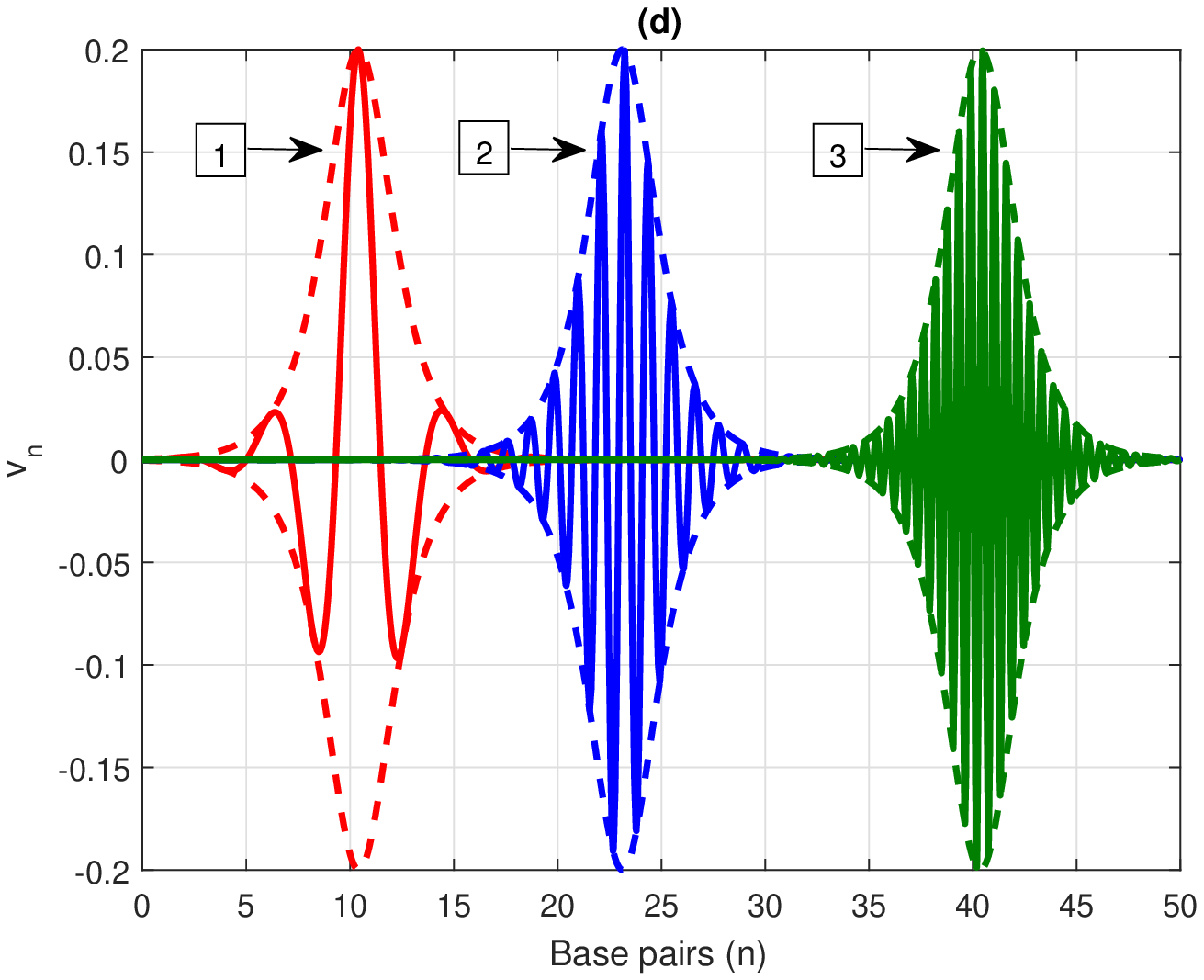}
\caption{Stretching of the nucleotide pair vs base pairs at $t=18$ $t.u.$, depending on the type of the inhomogeneity in the lattice, for $\varepsilon=0.1$ $c_0=0.18$, $\eta=1$, $q=\frac{\pi}{8}$ and $A=1$. The red lines (bubble (1)) represent  the bubble propagating in the homogeneous DNA chain, while the blue lines and the green lines (bubble (2) and bubble (3)) represent the bubbles propagating in the DNA chain in the presence of localized and exponential inhomogeneities, respectively. Dashed lines represent the envelope, while the solid lines represent the stretching of the base pairs}\label{fig4}
\end{figure}

\begin{figure}[H]
\centering
\includegraphics[width=3in]{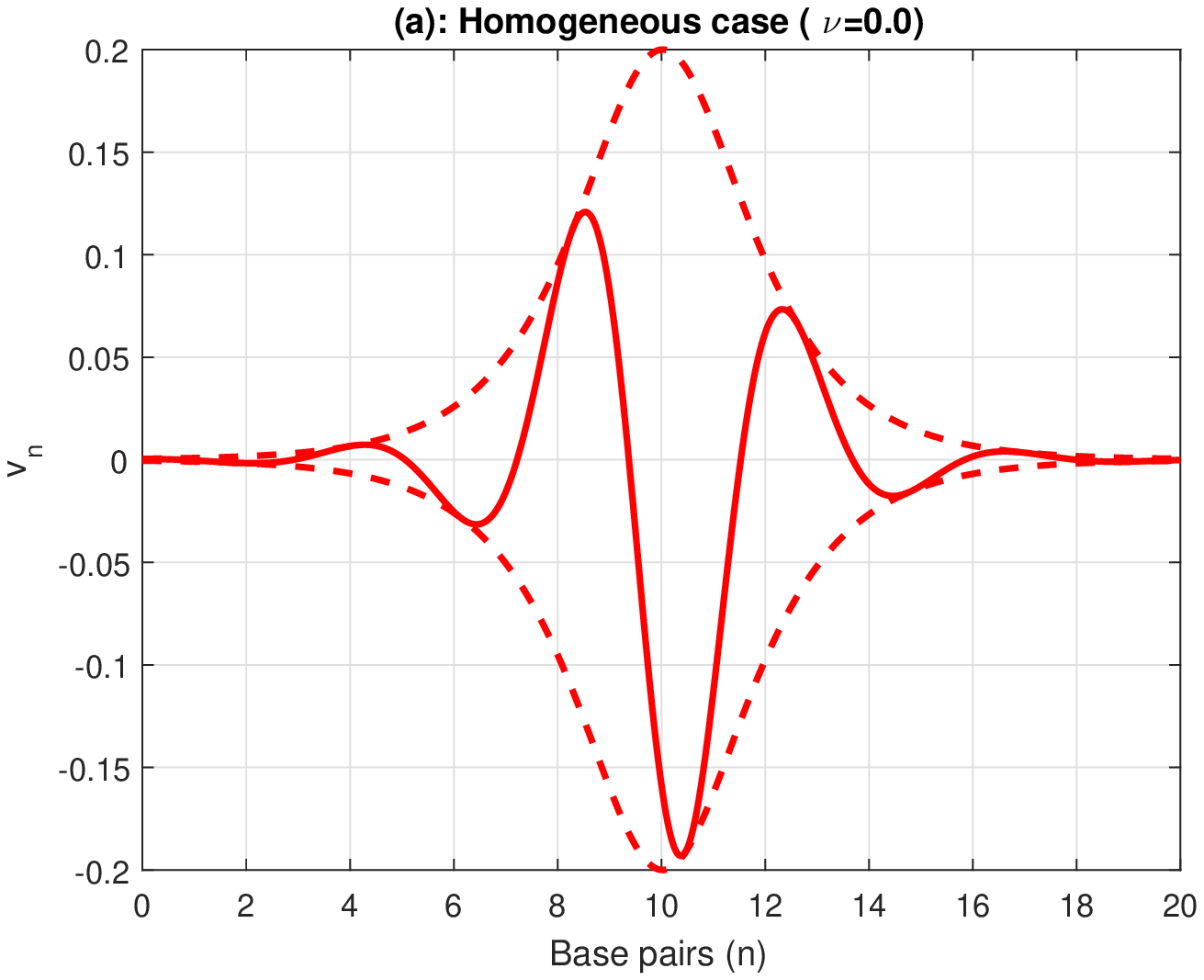}
\includegraphics[width=3in]{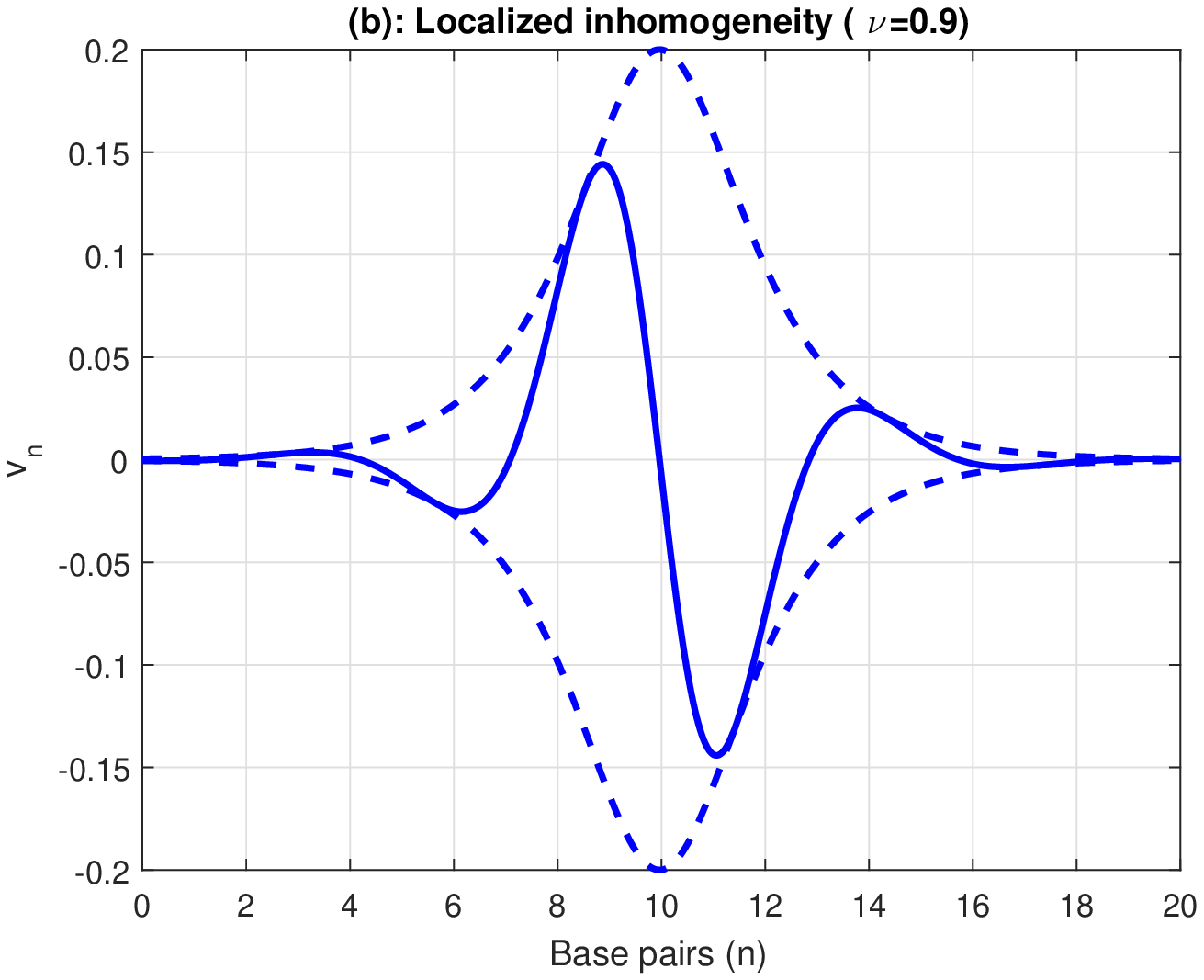}
\includegraphics[width=3in]{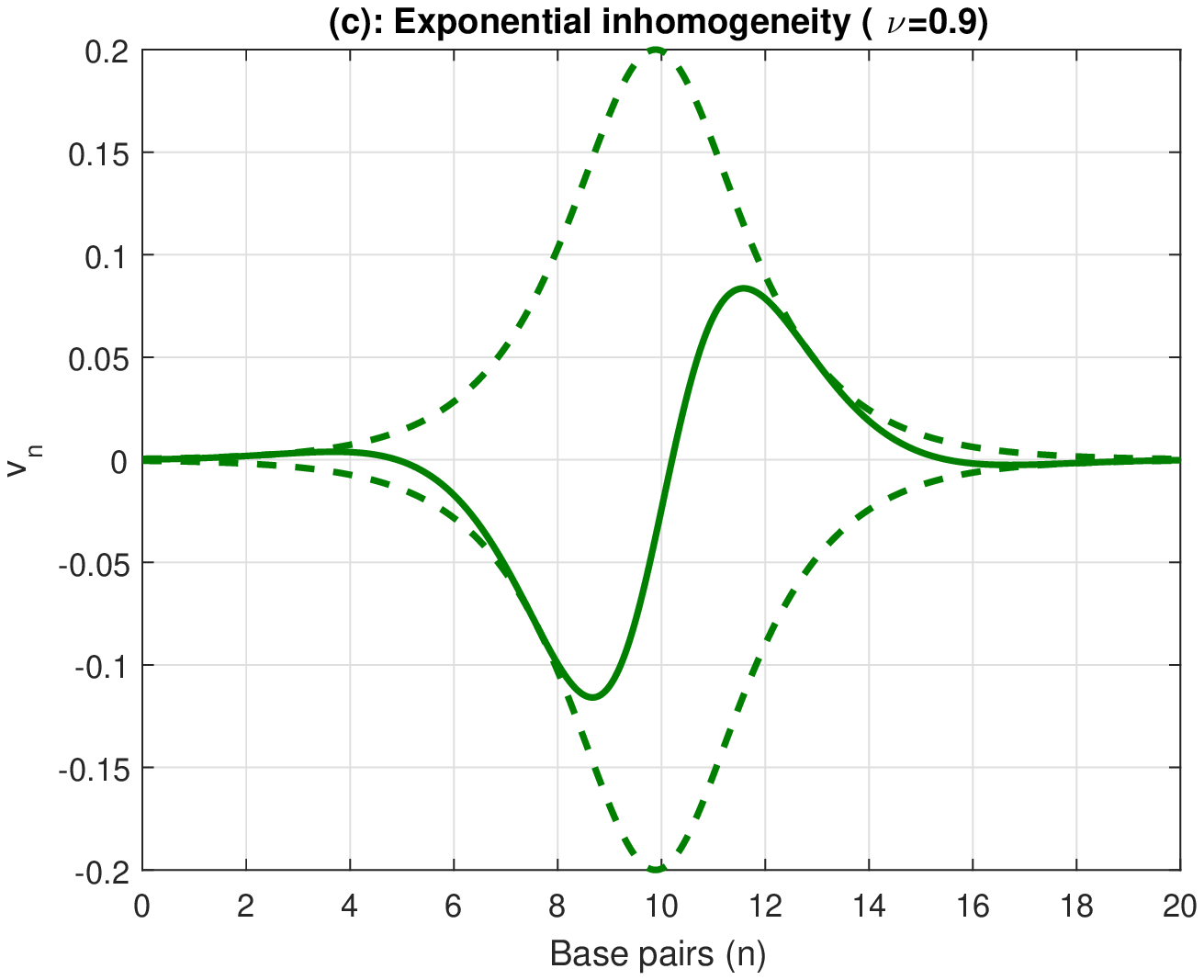}
\caption{Stretching of the nucleotide pair vs base pairs at $t=0.9$ $t.u.$, depending on the type of the inhomogeneity in the lattice, for $A=-1$. The other parameters are the same as in \figref{fig4}. .}\label{fig5}
\end{figure}

\begin{figure}[H]
\centering
\includegraphics[width=3in]{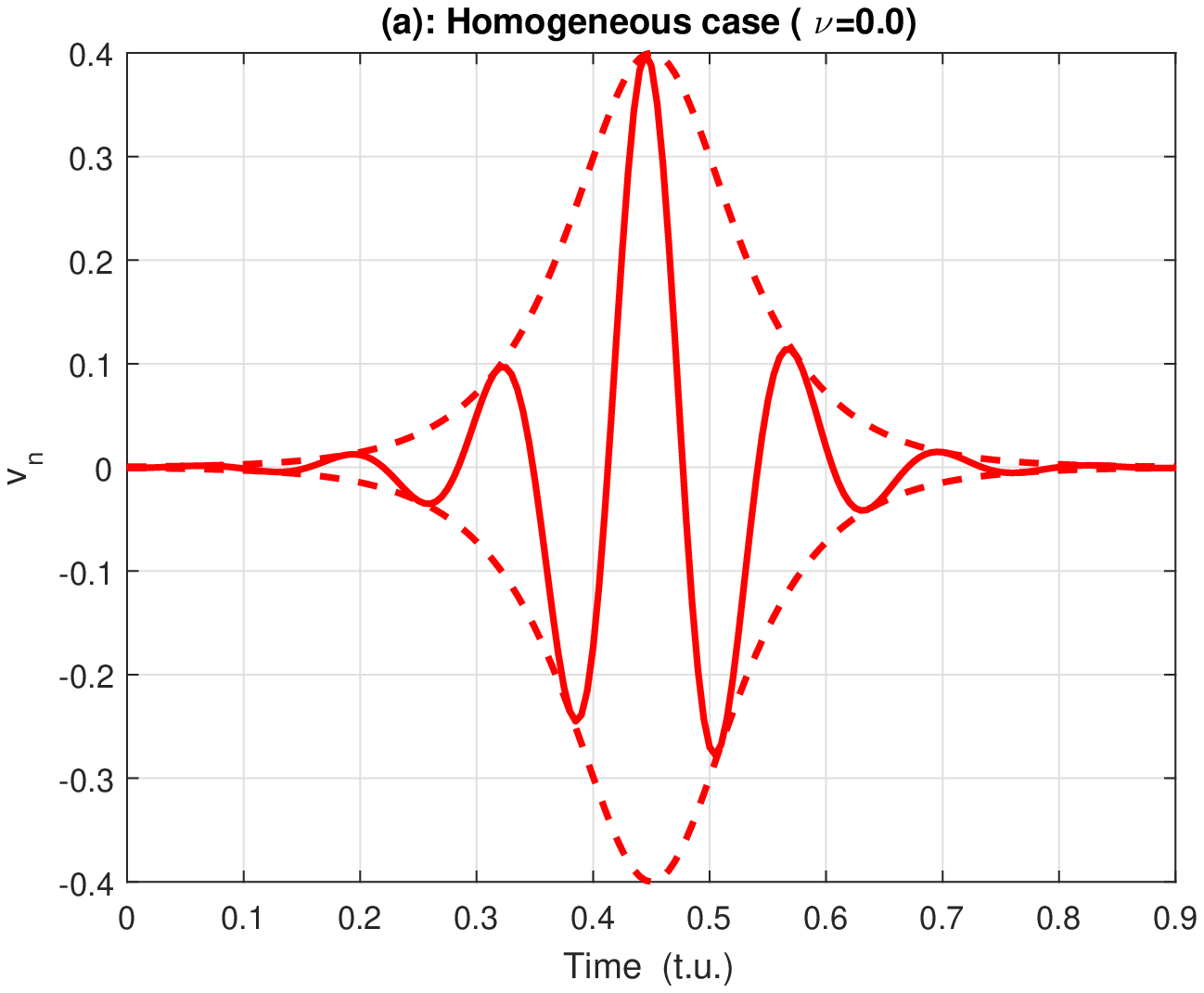}
\includegraphics[width=3in]{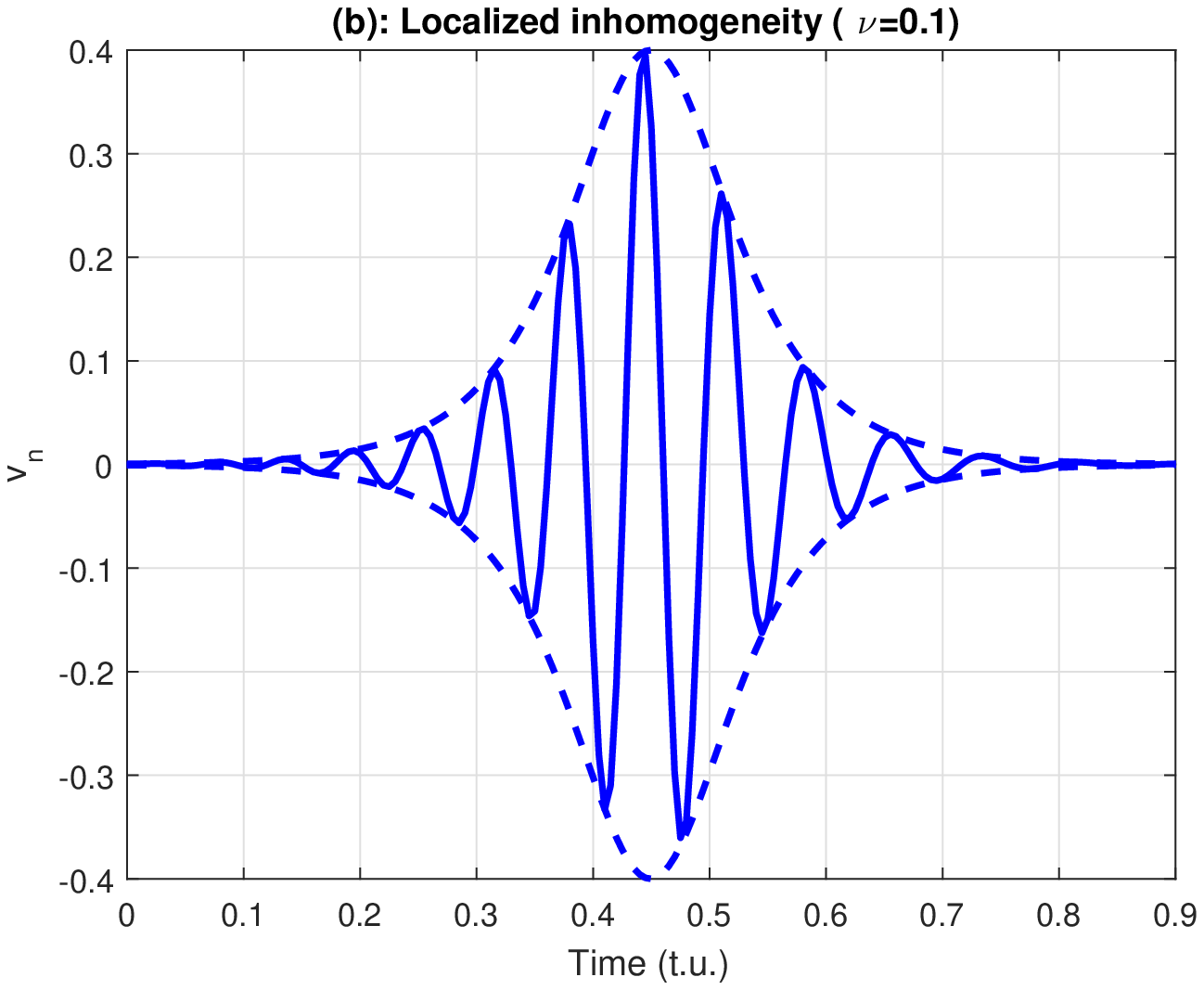}
\includegraphics[width=3in]{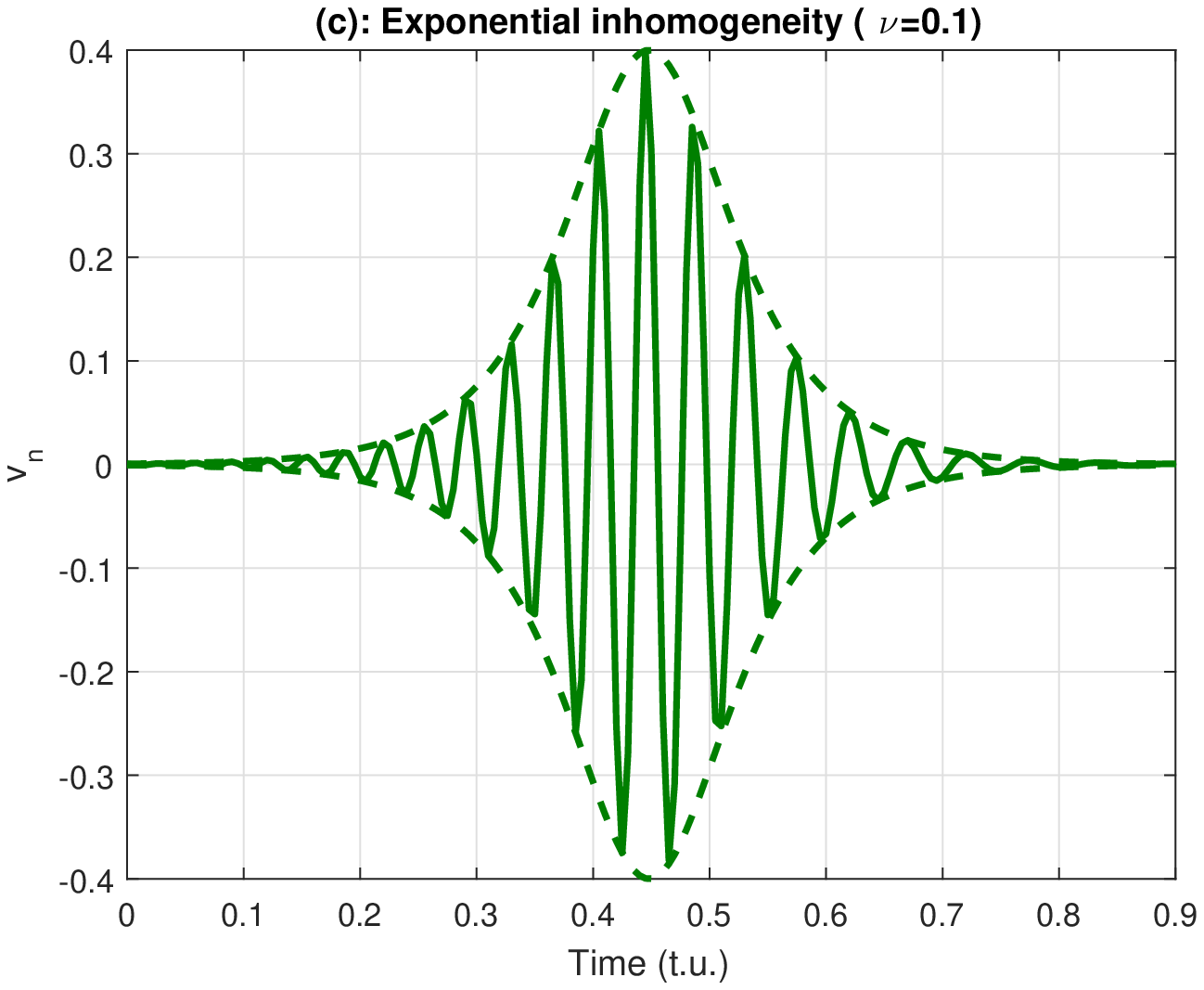}
\caption{Elongation of the out-of-phase motion vs time, depending on the type of the inhomogeneity in the lattice, for $\varepsilon=0.2$ $c_0=50$, $\eta=1$, $q=\frac{\pi}{8}$, $n=0$ and $A=1$. Panel (a) represents  the homogeneous case, panel (b) represents the case of localized inhomogeneity, while panel (c) represents the case of exponential inhomogeneity. Dashed lines represent the envelope, while the solid lines represent the stretching of the base pairs.}\label{fig6}
\end{figure}

\begin{figure}[H]
\centering
\includegraphics[width=3in]{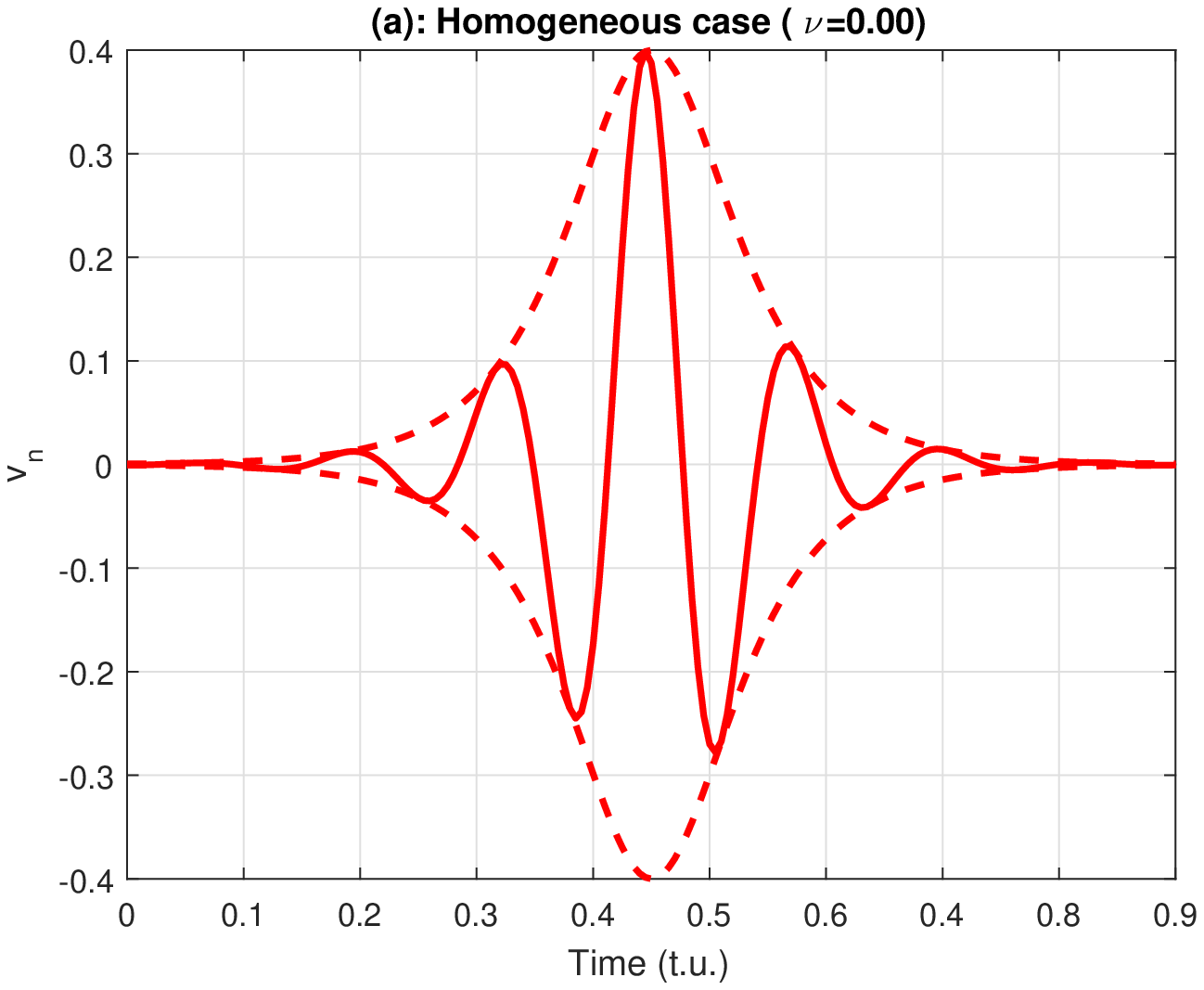}
\includegraphics[width=3in]{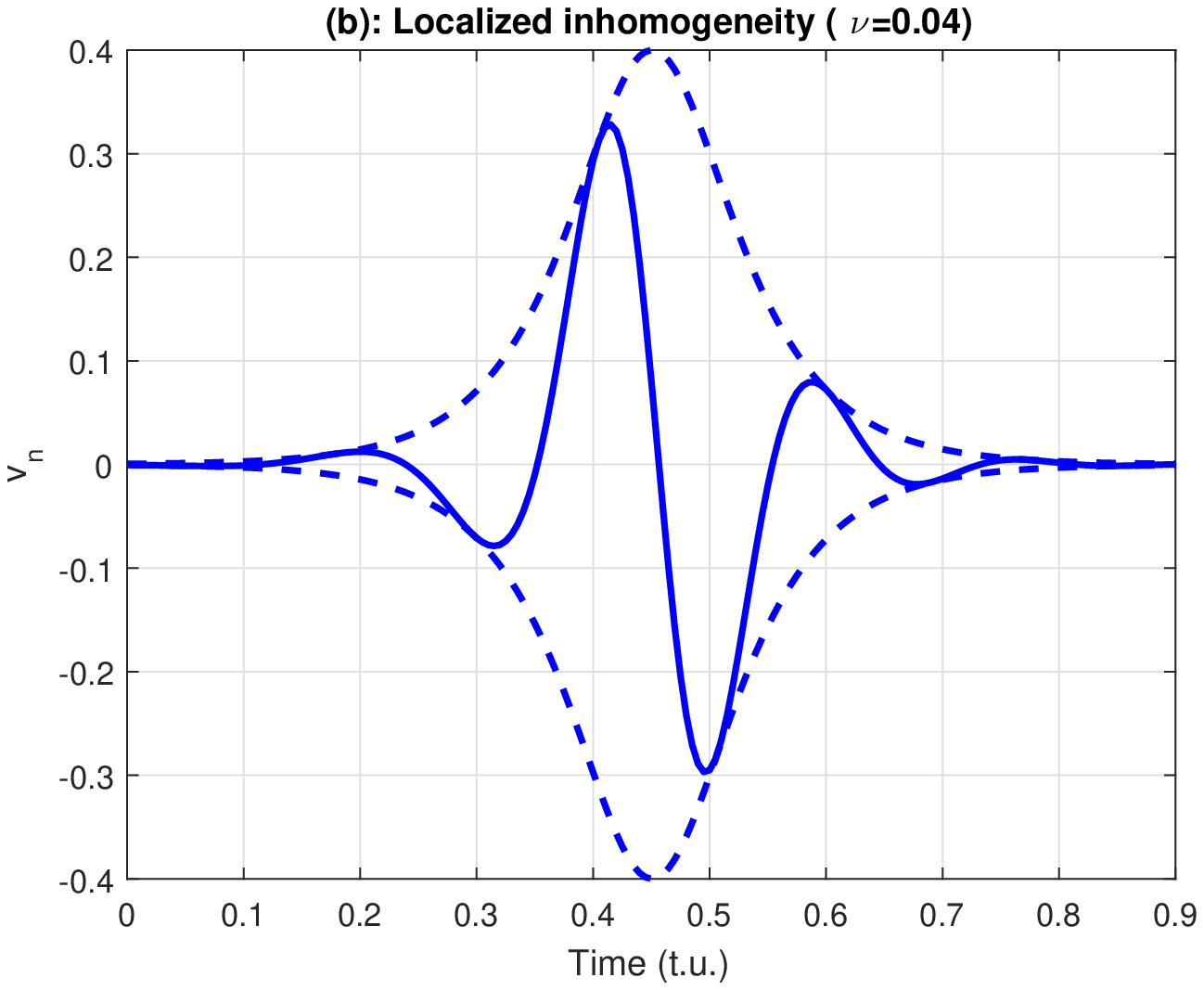}
\includegraphics[width=3in]{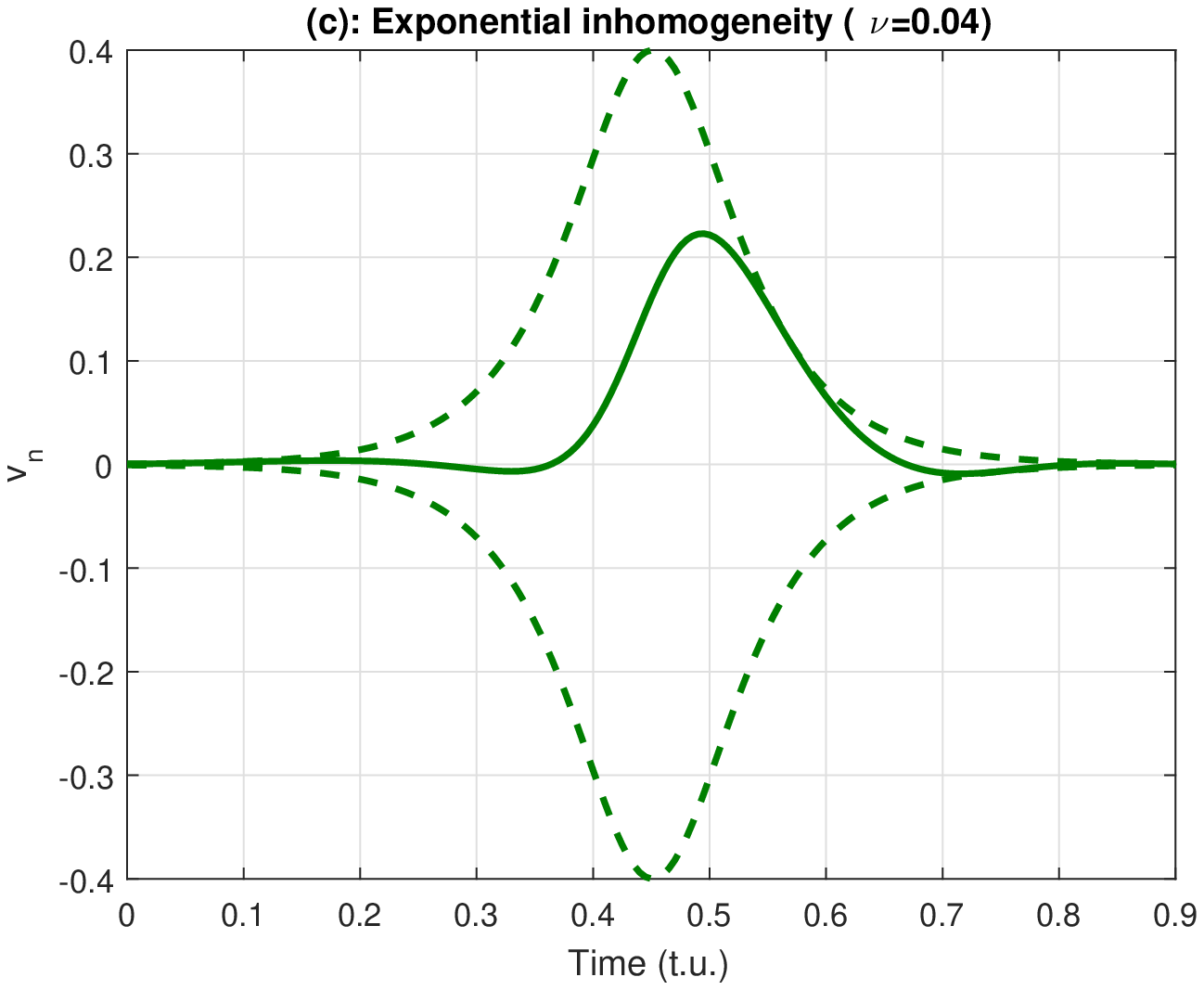}
\caption{Elongation of the out-of-phase motion vs time, depending on the type of the inhomogeneity in the lattice for $A=-1$. The other parameters are the same as in \figref{fig6}.}\label{fig7}
\end{figure}

The configuration of the molecule and the elongation of the out-of-phase motion are depicted as functions of the time, depending on the type and nature of the inhomogeneities present in the lattice. Figures \ref{fig4} and \ref{fig5} represent the configuration of the molecule, while Figures \ref{fig6} and \ref{fig7} present the elongation of the out-of-phase motion.
As predicted by Eqs. (\ref{eq42}), (\ref{eq43}) and (\ref{eq57}), we observe that the wave velocity, the wavenumber and the angular frequency of the soliton solution get a correction which can be positive or negative depending on the nature of the inhomogeneities:
\begin{description}
  \item  For $A>0$, the inhomogeneity behaves as an energetic barrier. The above parameters increase as time increases due to the inhomogeneities. \figref{fig4} shows the increasing of the wavenumber and wave velocity (\figref{fig4}d), while \figref{fig6} shows the increasing of the angular frequency. The increasing of the velocity helps the soliton to overcome this energetic barrier. Therefore, the soliton can propagate easily without formation of a bound state.
  \item For $A<0$, the decreasing of the above wave parameters is observed in \figref{fig5} for the wavenumber, and in \figref{fig7} for the angular frequency. In this case, the inhomogeneity behaves as a potential well in which the soliton is trapped. Thus, the soliton slows down, stops and vanishes at $t_0=\frac{-c_0}{B_i(\nu A)\eta}$, where $B_i$ is equal to $2/3$ and $\pi/2$ for localized and exponential inhomogeneities, respectively.
  \item For $A=0$, the inhomogeneity is switched off. This case corresponds to the soliton travelling in a homogeneous media with the constant velocity, wavenumber and angular frequency.
\end{description}

Also, as in Figures \ref{fig1} and \ref{fig2}, we observe that the absolute value of the corrections in the velocity and wavenumber (see \figref{fig4} and \figref{fig5}), and in the angular frequency (see \figref{fig6} and \figref{fig7}) of the soliton is greater in the case of exponential inhomogeneity. We also notice that the inhomogeneities do not affect the bubble sizes. The height and the width of the bubble remain constant during its propagation in the inhomogeneous DNA molecule.

It is clearly known that the transcription process is a very biological complex phenomenon \cite{yang}. Thus, the increasing of the velocity, the wavenumber and the angular frequency of the soliton during the crossing of the inhomogeneities makes difficult the execution of the biological functions of DNA such as reading, transcription and recombination of the genetic code to mention a few. These difficulties undoubtedly involve errors which can rise to mutations, damaging, missing of genes and biological diseases such as: sickle cell anemia, heart diseases, high blood pressure, Alzheimer's disease, diabetes, cancer, obesity, eye disease, epilepsy or stroke-like episodes which are due to mutations or disorder caused by a combination of environmental factors and mutations.
\section{\label{conclusion}Conclusion}

The wave dynamics of the PB model of DNA at the physiological temperature was studied. The classical PB model assumes the bases to be equal. In our model, we include the differences of bases and sequences through site dependent spring-constant. Using the multiple-scale expansion method in the semi-discrete approximation, the out-of-phase motion has been described by the Perturbed NLS equation. In the homogeneous limit, the dynamics is governed by the breather soliton of the integrable NLS equation. Indeed, to understand the effect of inhomogeneities on the base pairs opening, we carried out a perturbation analysis using multiple-scale soliton perturbation technique. For implement this, we linearized the Perturbed NLS equation with the aids of Poincar\'{e}-type asymptotic expansion. The breatherlike soliton solution which represents the opening of base pairs travelling along the inhomogeneous DNA chain in the form of bubble has been constructed for different forms of the inhomogeneities (localized and exponential).

The results showed that the bubble profile is not affected by the inhomogeneities. However, the velocity with which the base pairs are opening or closing, the wavenumber with which the soliton propagates and angular frequency with which the base pairs vibrate increase, decrease or remain uniform and even the soliton stops depending on the nature of the inhomogeneities. The high velocity, wavenumber and angular frequency of the soliton can create disorder in the execution of DNA biological functions and make coding or reading errors which can leads to varieties of biological diseases.

The inhomogeneous DNA models can explain DNA biological functions such as replication, transcription and recombination more viably than the homogeneous one, since it can explain and predict the biological diseases due to genetic mutations. The number of DNA damages in a single human cell exceeds $10,000$ every day \cite{klu}, and must be counteracted by special DNA repair processes. Long-range interactions (LRI) are ubiquitous in DNA molecule, they play a crucial role in the stabilization of the molecule \cite{oka}. Hence, it is important to understand the interplay between inhomogeneity and LRI \cite{saha2}. These studies are now in progress. Despite the relevance of this work, the impact of inhomogeneities on the DNA biological processes is not yet clearly understood, since the nature generally selects inhomogeneous DNA and the gap between the nonlinear physics of DNA and medicine is still big.
\\
\\
\acknowledgement{J. B. Okaly is in debt to Dr. Ndzana Fabien II of the University of Maroua, Maroua- Cameroon for some recommendations and fruitful discussions.}
 \newpage
%m=300; D=0.03; r=3.4; a=4.45; K=0.06; K1=K/m; epsilon=0.1; q=pi/8;  N=20; c0=0.18; eta=1; nu=0.1; A=-1;
% c0=0.01;  eta=0.1;   nu=0.01; mu=0; A=0.1;   N=50;q=pi/8; epsilon=0.5;
%m=300; D=0.03; r=3.4; a=4.45; K=0.06; K1=K/m; epsilon=0.1; q=pi/8;  N=20; c0=0.18; eta=1; nu=0.0; A=-1; t=0.9;
%%epsilon=0.2; q=pi/8;  N=20;c0=50; eta=1; nu=0.1; A=1;n=0;
%m=300; D=0.03; r=3.4; a=4.45; K=0.06; K1=K/m; epsilon=0.2; q=pi/8;  N=20;c0=50; eta=1; nu=0.04; A=-1;n=0;

\begin{thebibliography}{apssamp}
%{99}
\section*{References}
%
\bibitem{sty} L. Styer, \emph{Biochemistry, fourth ed.}, W. H. Freeman and Company, New York, 1995.
%
\bibitem{kal} G. Kalosakas, K. Q. Rasmussen and A. R. Bishop, \emph{Nonlinear excitations in DNA: polarons and bubbles}, Synth. Met. \textbf{141} (2004) 93-97.
%
\bibitem{Englander} S. W. Englander, N. R. Kallenbach, A. J. Heeger, J. A. Krumhansl and S. Litwint, \emph{\emph{Nature of the Open State in Long Polynucleotide Double Helices: Possibility of Soliton Excitations}}, Proc. Natl. Acad. Sci. \textbf{77} (1980) 7222-7226.
%
\bibitem{yak1} L. V. Yakushevich, \emph{Nonlinear DNA dynamics: hierarchy of the models}, Physica D \textbf{79} (1994) 77-86.
%
\bibitem{yak2} L. V. Yakushevich, \emph{Nonlinear DNA dynamics: a new model}, Phys. Lett. A \textbf{136} (1989) 413-417.
%
\bibitem{kong} K. De-Xing, L. Sen-Yue and Z. Jin, \emph{Nonlinear dynamics in a new double chain-model of DNA}, Commun. Theor. Phys. \textbf{36} (2001) 737-742.
%
    \bibitem{oka2} J. B. Okaly, A. Mvogo, R. L. Woulach\'{e} and T. C. Kofan\'{e}, \emph{Semi-discrete Breather in a Helicoidal DNA Double Chain-Model} (Unpublished results).
%
\bibitem{PB} M. Peyrard and A. R. Bishop, \emph{Statistical Mechanics of a Nonlinear Model for DNA Denaturation}, Phys. Rev. Lett. \textbf{62} (1989) 2755-2758.
%
\bibitem{dau} T. Dauxois, \emph{Dynamics of breather modes in a nonlinear \textquotedblleft helicoidal\textquotedblright\; model of DNA}, Phys. Lett. A. \textbf{159} (1991) 390-395.
    %
\bibitem{lad} J. Ladik, J. Cizek, \emph{Probable physical mechanisms of the activation of oncogenes through carcinogens}, Int. J. Quantum Chem. \textbf{26} (1984) 955-964.
%
\bibitem{cube} E. Cubero, E. C. Sherer, F. J. Luque, M. Orozco, C.A. Laughton, \emph{Observation of Spontaneous Base Pair Breathing Events in the Molecular Dynamics Simulation of a Difluorotoluene-Containing DNA Oligonucleotide}, J. Am. Chem. Soc. \textbf{121} (1999) 8653-8654.
%
\bibitem{danD} M. Daniel and V. Vasumathi, \emph{Perturbed soliton excitations in the DNA double helix}, Physica D \textbf{231} (2007) 10-29.
%
\bibitem{danA} M. Daniel and V. Vasumathi, \emph{Nonlinear molecular excitations in a completely inhomogeneous DNA chain}, Phys. Lett. A \textbf{372} (2008) 5144-5151.
%
\bibitem{agu} M. A. Ag\"uero , T. L. Belyaeva  and V. N. Serkin, \emph{Compacton anti-compacton pair for hydrogen bonds and rotational waves in DNA dynamics}, Commun. Nonlinear Sci. Numer. Simulat. \textbf{16} (2011) 3071-3080.
%
\bibitem{val} M. Valko, C. J. Rhodes, J. Moncol, M. Izakovic and M. Mazur, \emph{Free radicals, metals and antioxidants in oxidative stress-induced cancer}, Chem. Biol. Interact. \textbf{160} (2006) 1-40.
%
\bibitem{kawa} S. Kawanishi, Y. Hiraku, S. Pinlaor and N. Ma, \emph{Oxidative and nitrative DNA damage in animals and patients with inflammatory diseases in relation to inflammation-related carcinogenesis}, Biol. Chem. \textbf{387} (2006) 365-372.
%
\bibitem{khan} K. K. Khanna, S. P. Jackson, \emph{DNA double-strand breaks: signaling, repair and the cancer connection}, Nature Genet. \textbf{27} (2001) 247-254.
%
\bibitem{jack} S. P. Jackson and R. J. Bartek, \emph{The DNA-damage response in human biology and disease}, Nature Rev. \textbf{461} (2009) 1071-1078.
%
\bibitem{bie} J. H. Bielas, K. R. Loeb, B. P. Rubin, L. D. True and Loeb, \emph{Human cancers express a mutator phenotype}, Proc. Natl Acad. Sci. \textbf{103} (2006) 18238-18242.
%
\bibitem{yaku1}  L. V. Yakushevich, \emph{Nonlinear Physics of DNA}, Wiley, Chichester, 2004.
%
\bibitem{kyl} K. Forinash, M. Peyrard and B. Malomed, \emph{\emph{Interaction of discrete breathers with impurity modes}}, Phys. Rev. E \textbf{49} (1994) 3400-3411.
%
\bibitem{gra} W. Gratzer \emph{Association of nucleic-acid bases in aqueous solution: A solvent partition study}. Eur. J. Biochem. \textbf{10} (1969) 184-187.
%
\bibitem{davi} D. B. Davies \emph{Co-operative conformational properties of nucleosides, nucleotides and nucleotidyl units in solution}, B. Pullmlln (ed.), Nuclear Magnetic Resonance Spectroscopy in Molecular Biology, D. Reidel Publishing Company, Dordrecht, Holland, 1978, pp. 71-85.
%
\bibitem{orns} R. L. Ornstein, R. Rein, D. L. Breen and R. D. MacElroy \emph{An optimized potential function for calculation of nucleic acid interaction energies. I. Base stacking}. Biopolymers \textbf{17} (1978) 2341-2360.
%
\bibitem{oka} J. B. Okaly, A. Mvogo, R. L. Woulach\'{e} and T. C. Kofan\'{e}, \emph{Nonlinear dynamics of damped DNA systems with long-range interactions}, Commun. Nonlinear Sci. Numer. Simulat., \textbf{55} (2018) 183-193.
%
\bibitem{pey} M. Peyrard, \emph{Nonlinear dynamics and statistical physics of DNA}, Nonlinearity \textbf{17} (2004) R1-R40.
%
\bibitem{zdr} S. Zdravkovi\'c and S. Zekovi\'c, \emph{Nonlinear dynamics of microtubules and series expansion unknown function method}, Chinese. J. Phys. \textbf{55} (2017) 2400-2406.
%
\bibitem{tala} E. Tala-Tebue, Z. I. Djoufack, D .C. Tsobgni-Fozap, A. Kenfack-Jiotsa, F. Kapche-Tagne, T. C. Kofané, \emph{Traveling wave solutions along microtubules and in the Zhiber-Shabat equation}, Chinese J. Phys., \textbf{55} (2017) 939-946.
%
\bibitem{rem} M. Remoissenet, \emph{Low-amplitude breather and envelope solitons in quasi-one-dimensional physical models}, Phys. Rev. B  \textbf{33} (1986) 2386-2392.
%
\bibitem{jia} J. Yan, Y. Tang, and G. Zhou, \emph{Direct approach to the study of soliton perturbations of the nonlinear Schr\"{o}dinger equation and the sine-Gordon equation}, Phys. Rev. E \textbf{58} (1998) 1064-1073.
%
\bibitem{danE} V. Vasumathi and M. Daniel, \emph{Base-pair opening and bubble transport in a DNA double helix induced by a protein molecule in a viscous medium}, Phys. Rev. E \textbf{80} (1-9) (2009), 061904.
%
\bibitem{karp} V. I. Karpman, \emph{Soliton evolution in the presence of perturbation}, Phys. Scr. \textbf{20} (1979) 462-478.
%
\bibitem{karp2} V. I. Karpman and V. V. Solov'ev, \emph{A perturbation theory for soliton systems}, Physica D \textbf{3} (1981) 142-164.
%
\bibitem{mvo} A. Mvogo, G.H. Ben-Bolie and T. C. Kofan\'{e}, \emph{Solitary waves in an inhomogeneous chain of $\alpha$-helical proteins}, Int. J. of Mod. Phys B \textbf{28} (1-14) (2014), 1450109.
%
\bibitem{saha} M. Saha and T. C. Kofan\'e, \emph{inhomogeneities and nonlinear dynamics of a helical DNA interacting with a RNA-polymerase}, Phys. Scr. \textbf{89} (1-7) (2014), 085003.
%
\bibitem{putn} B. F. Putnam, L. L. Van Zandt, E. W. Prohofsky, M. N. Mei, \emph{Resonant and localized breathing modes in terminal regions of the DNA double helix}, Biophys J. \textbf{35} (1981) 271-287.
%
\bibitem{proh} E. W. Prohofsky, K. C. Lu, L. L. Van Zandt and B.F. Putnam, \emph{Breathing modes and induced resonant melting of the double helix},  Phys. Lett. A \textbf{70} (1979) 492-494.
%
\bibitem{yang} X. Yang, Y. Wu a, Z. Yuan, \emph{Characteristics of mRNA dynamics in a multi-on model of stochastic transcription with regulation}, Chinese J. Phys. \textbf{55} (2017) 508-518.
%
\bibitem{klu} A. Klungland, Y-G Yang, \emph{Endogenous DNA Damage and Repair Enzymes: -A short summary of the scientific achievements of Tomas Lindahl, Nobel Laureate in Chemistry 2015}, Genomics Proteomics Bioinformatics \textbf{14} (2016) 122-125.
%
\bibitem{saha2} M. Saha and T. C. Kofan\'e, \emph{Long-range interactions between adjacent and distant bases in a DNA and their impact on the ribonucleic acid polymerase-DNA dynamics}, Chaos \textbf{22} (1-12) (2012), 013116.
%
\end{thebibliography}
\end{document}
%\endinput